\documentclass[fleqn,usenatbib,useAMS]{mnras}

\pdfoutput=1
\usepackage{hyperref}
\usepackage{xcolor}

\usepackage{color}
\usepackage{float}
\usepackage{subfigure}
\usepackage{tabularx}
\usepackage{soul}
\usepackage{comment}

\usepackage{graphicx}	
\usepackage{amsmath}	
\usepackage{amssymb}	
\usepackage{multicol}        
\usepackage{amsfonts}
\usepackage{pdflscape}	

\usepackage[T1]{fontenc}
\usepackage{ae,aecompl}

\usepackage{newtxtext,newtxmath}

\usepackage{graphicx}
\usepackage{dcolumn}
\usepackage{amsmath}
\usepackage{amsfonts}
\usepackage{amssymb}
\usepackage{color}
\usepackage{float}
\usepackage{subfigure}
\usepackage{tabularx}
\usepackage{soul}

\newcolumntype{L}[1]{>{\raggedright\arraybackslash}p{#1}}
\newcolumntype{C}[1]{>{\centering\arraybackslash}p{#1}}
\newcolumntype{R}[1]{>{\raggedleft\arraybackslash}p{#1}}

\definecolor{v}{rgb}{0.6, 0.2, 0.8} 

\providecommand{\U}[1]{\protect\rule{.1in}{.1in}}

\usepackage{tikz,xcolor,hyperref}

\definecolor{lime}{HTML}{A6CE39}
\DeclareRobustCommand{\orcidicon}{%
	\begin{tikzpicture}
	\draw[lime, fill=lime] (0,0) 
	circle [radius=0.16] 
	node[white] {{\fontfamily{qag}\selectfont \tiny ID}};
	\draw[white, fill=white] (-0.0625,0.095) 
	circle [radius=0.007];
	\end{tikzpicture}
	\hspace{-2mm}
}

\foreach \x in {A, ..., Z}{%
	\expandafter\xdef\csname orcid\x\endcsname{\noexpand\href{https://orcid.org/\csname orcidauthor\x\endcsname}{\noexpand\orcidicon}}
}

\title{Observational constraints and dynamical analysis of 
Kaniadakis horizon-entropy cosmology}

\author[A.  Hern\'andez-Almada, et. al.]{ A.  Hern\'andez-Almada\orcidA{}$^{1}$ 
\thanks{Contact e-mail: \href{mailto:ahalmada@uaq.mx}{ahalmada@uaq.mx}},  Genly Leon\orcidB{}$^{2,3}$, Juan Maga\~na\orcidC{}$^4$,  Miguel A. Garc\'ia-Aspeitia\orcidD{}$^{5}$,  
\newauthor V. Motta\orcidE{}$^6$, Emmanuel N. Saridakis\orcidF{}$^{7,8,9}$,  Kuralay Yesmakhanova\orcidG{}$^{9,10}$, Alfredo D. Millano\orcidH{}$^2$
\\
$^{1}$Facultad de Ingenier\'ia, Universidad Aut\'onoma de 
Quer\'etaro, Centro Universitario Cerro de las Campanas, 76010, Santiago de 
Quer\'etaro, M\'exico,\\
$^2$Departamento  de  Matem\'aticas,  Universidad Cat\'olica del Norte, Avda.   Angamos  0610,  Casilla  1280  Antofagasta,  Chile,\\
$^3$Institute of Systems Science, Durban University of Technology, PO Box 1334,
Durban 4000, South Africa\\
$^4$Instituto de Astrof\'isica \& Centro de Astro-Ingenier\'ia, Pontificia Universidad Cat\'olica de Chile, Av. Vicu\~na Mackenna, 4860, Santiago, Chile,\\
$^5$ Depto. de Física y Matemáticas, Universidad Iberoamericana Ciudad de México, Prolongación Paseo \\ de la Reforma 880, México D. F. 01219, México \\
$^6$Instituto de F\'isica y Astronom\'ia, Facultad de Ciencias,  Universidad de Valpara\'iso, \\ Avda. Gran Breta\~na 1111, Valpara\'iso, Chile,\\
$^7$Institute for Astronomy, Astrophysics, Space Applications and Remote Sensing, National Observatory of Athens, Lofos Nymfon, 11852 Athens,  Greece,\\
$^8$CAS Key Laboratory for Researches in Galaxies and Cosmology,  Department of Astronomy, \\ University of Science and Technology of China, Hefei, Anhui 230026, P.R. China,\\
$^{9}$Ratbay Myrzakulov Eurasian International Centre for Theoretical Physics, Nur-Sultan 010009,  Kazakhstan,\\
$^{10}$Ratbay Myrzakulov Eurasian International Centre for Theoretical Physics, Eurasian National University, Nur-Sultan Astana 010008, Kazakhstan.
}

\date{\today}

\pubyear{2022}

\begin{document}
\label{firstpage}
\pagerange{\pageref{firstpage}--\pageref{lastpage}}
\maketitle

\begin{abstract}
We study the scenario of   Kanadiakis horizon entropy cosmology 
which arises  from the application of the gravity-thermodynamics conjecture 
using the Kaniadakis modified entropy. The resulting  modified Friedmann 
equations   contain extra terms that constitute an effective dark energy 
sector. We use data from   Cosmic chronometers, Supernova Type Ia,  
HII galaxies, Strong lensing systems,  and  Baryon acoustic oscillations observations and we apply  a Bayesian Markov Chain Monte Carlo analysis to construct the likelihood contours 
for the model parameters. We find that the Kaniadakis parameter  
 is constrained  around 0, namely,  around the value 
 where the standard 
Bekenstein-Hawking is recovered.   
Concerning the normalized 
Hubble parameter, we find  $h=0.708^{+0.012}_{-0.011}$, a result  that is 
independently verified by applying   the  $\mathbf{\mathbb{H}}0(z)$ 
diagnostic and, thus, we conclude that the scenario at hand 
can alleviate the  $H_0$ tension problem.  
Regarding the transition redshift,  
 the reconstruction of the cosmographic parameters
gives $z_T=0.715^{+0.042}_{-0.041}$. Furthermore, 
we apply the AICc,  BIC and DIC information criteria and we find that in most 
datasets the scenario is  statistical equivalent to $\Lambda$CDM one. Moreover, 
we examine the Big Bang Nucleosynthesis (BBN) and we show that the scenario  
satisfies the corresponding requirements. 
Additionally, we perform a   phase-space  analysis,  and we show that the 
Universe 
past attractor  is the 
matter-dominated epoch, while at  late times the Universe results in the 
dark-energy-dominated solution.  Finally, we show that  Kanadiakis 
horizon entropy cosmology  accepts heteroclinic sequences, but it cannot exhibit bounce and 
turnaround solutions.
\end{abstract}

\begin{keywords}
theory, dark energy, cosmological parameters.
\end{keywords}

\date{\today}
\maketitle
\flushbottom
\section{Introduction}

Cosmology is one of the most exciting adventures in the human endeavour: the understanding of the origin, evolution and future of our Universe by combining the physics at micro- and macro-scales in a joint framework. In particular, the Universe acceleration is one of the most important puzzles. Discovered by the Supernovae team, through type Ia supernovae \citep[SNIa,][]{Riess:1998}, it is also confirmed by the acoustic peaks of the Cosmic Microwave Background Radiation
\citep[CMB,][]{WMAP:2003elm}, and recently tested with large-scale structure measurements \citep{Nadathur:2020kvq}. Tackling the Universe acceleration is indeed a complicated issue due to the  attractive nature of gravity in the General Relativity (GR) framework.
The first approach on its explanation is through the addition of  the cosmological constant 
\citep[CC,][]{Carroll:2000}. Thus, using the continuity equation   and assuming a constant energy density ($\dot{\rho}=0$), it is possible to conclude that the equation of state (EoS) is $w=-1$, which is in concordance with what is expected for a fluid that accelerates the Universe. Another main characteristic is that the energy density of the CC must be subdominant in order to obtain a late and non-violent acceleration. From the Quantum Field Theory (QFT) viewpoint, the CC can be explained by the addition of 
quantum vacuum fluctuations (QVF) associated to the space-time. Therefore, the expansion of space-time implies an increase of QVF, maintaining a constant energy density. However, when we calculate the energy density of the QVF, the result is in complete disagreement with the observed one \citep[see][]{Zeldovich:1968ehl,Weinberg}, this is the so-called {\it fine tuning problem} \citep{Addazi:2021xuf}.
In addition to the cosmological constant problem, a $4.2\sigma$ tension in the current Hubble parameter value ($H_0$) measured by Supernova $H_0$ for the Equation of State (SH0ES) collaboration \citep{Riess:2020fzl} and the one obtained by Planck collaboration under the $\Lambda$-Cold Dark Matter ($\Lambda$CDM) scenario  \citep{Planck:2018vyg} has been recently observed.

The above mentioned problems that afflicts the understanding of the CC in this framework has driven the community to propose other approaches like scalar fields, dynamical dark energy, viscous fluids, Chaplygin gas \citep{Hernandez-Almada:2018osh} or modifications to GR  such as unimodular gravity \citep{Garcia-Aspeitia:2019yni,Garcia-Aspeitia:2019yod}, Einstein-Gauss-Bonet \citep{Garcia-Aspeitia:2020uwq}, Brane Worlds \citep{Garcia-Aspeitia:2018fvw}, among others 
\citep[see][for a compilation of the mentioned previous models]{CANTATA:2021ktz,Motta:2021hvl}. 
Despite the fact that we can have numerous models and scenarios describing the late time acceleration, at the end of the day the detailed confrontation with observations, alongside theoretical consistency, will be the main method for 
their validation.

One interesting  alternative to investigate the dynamics of the Universe is the gravity-thermodynamics approach \citep{Jacobson:1995ab,Padmanabhan:2003gd,Padmanabhan:2009vy}. This takes advantage of the first law of thermodynamics and the standard Bekenstein-Hawking entropy-area relation for black holes, which  applied to the apparent cosmological horizon leads to the Friedmann equations \citep{Frolov:2002va,Cai:2005ra,Akbar:2006kj,Cai:2006rs}. Hence, the application of this conjecture with  various alternative entropy relations  leads to modified Friedmann equations whose additional     terms can source the cosmic acceleration \citep{Cai:2006rs,Akbar:2006er,Paranjape:2006ca,Sheykhi:2007zp,Jamil:2009eb,Cai:2009ph,Wang:2009zv,Jamil:2010di,Sheykhi:2010wm,Sheykhi:2010zz,Gim:2014nba,Fan:2014ala,Lymperis:2018iuz,Sheykhi:2018dpn,Saridakis:2020lrg}.

One interesting class of extended entropies arises through
 generalizations for the Boltzmann-Gibbs statistics. In particular, one  modifies
 the classical entropy of a system  given by $S=-k_{B} \sum_{i}^{} p_{i} \ln{p_{i}}$, where $p_{i}$ is the probability of a system to be within a microstate, through non-extensive 
 analysis  
 resulting into  the Tsallis entropy \citep{Tsallis:1988, Lyra:1998}, through quantum-gravitational considerations resulting into Barrow entropy \citep{Barrow:2020tzx}, or through relativistic extensions resulting into Kaniadakis entropy \citep{Kaniadakis:2002zz,Kaniadakis:2005zk}.
 Hence, the application of the gravity-thermodynamics conjecture using the above modified horizon entropy gives rise to modified cosmological scenarios. In \citet{Lymperis:2018iuz} this was performed in the framework of Tsallis entropy, in 
 \citet{Saridakis:2020lrg,Saridakis:2020cqq,Leon:2021wyx,Barrow:2020kug} 
 applied it for the Barrow entropy case, and recently 
 \citep{Lymperis:2021qty} use it in the Kaniadakis entropy frame.
 
 In the present work, we 
 investigate Kaniadakis horizon entropy cosmology by confronting it to observational data from Cosmic Chronometers, Supernovae Type I (SNIa), HII galaxies, strong lensing systems (SLS), and baryon acoustic oscillations (BAO) observations. Additionally, we 
 perform a complete dynamical system analysis in order to etxract information of the local and global features of the cosmological evolution.

The outline of the paper is as follows: Sec. \ref{sec:cosmo} introduces the framework of the Kaniadakis cosmology. In Sec. \ref{sec:data} we infer the cosmological parameter under the Kaniadakis model using five observational datasets mentioned above. 
The section \ref{sec:SA}
presents a stability analysis around the equilibrium points of the dynamical system under the Kaniadakis cosmology. Finally, we discuss and summarize our results in Section \ref{sec:Con}. From now on we use natural units in which $\hbar=k_B=c=1$.


\section{Kaniadakis horizon entropy   cosmology} \label{sec:cosmo}

In this section we present the scenario of Kaniadakis horizon entropy   
cosmology  \citep{Lymperis:2021qty}, namely the modified Friedmann equations 
arising from the application of the gravity-thermodynamics conjecture using the 
extended Kaniadakis entropy.

Kaniadakis entropy  is an  one-parameter
generalization of the classical   entropy,  given as
$S_{K}=- k_{_B} \sum_i n_i\, \ln_{_{\{{\scriptstyle
K}\}}}\!n_i $ \citep{Kaniadakis:2002zz,Kaniadakis:2005zk},
with $k_{_B}$   the Boltzmann constant   and
where $\ln_{_{\{{\scriptstyle
K}\}}}\!x=(x^{K}-x^{-K})/2K$. In such a framework  the  
 dimensionless parameter $-1<K<1$ quantifies  the relativistic
deviations from  standard statistical mechanics, and the latter is recovered in 
the limit   $K\rightarrow0$.
  Kaniadakis entropy    can be expressed as 
\citep{Abreu:2016avj,Abreu:2017hiy,Abreu:2021avp}
\begin{equation}
 \label{kstat}
S_{K} =-k_{_B}\sum^{W}_{i=1}\frac{P^{1+K}_{i}-P^{1-K}_{i}}{2K},
\end{equation}
with   $P_i$    the probability of  a specific microstate  and 
 $W$  the total configuration number. When we apply it in the   
   black-hole framework,  we obtain
\citep{Drepanou:2021jiv,Moradpour:2020dfm,Lymperis:2021qty}
 \begin{equation} \label{kentropy}
S_{K} = \frac{1}{K}\sinh{(K S_{BH})},
\end{equation}
where 
\begin{equation}
\label{Horentropy}
S_{BH}=\frac{1}{4G} A,
\end{equation} 
is the   usual Bekenstein-Hawking entropy, with  $A$    the horizon 
area and $G$ is the gravitational constant.
Hence, in the limit  $K\rightarrow 0$  we  
recover Bekenstein-Hawking entropy.

Let is now apply the gravity-thermodynamics conjecture using   Kaniadakis 
entropy. We   consider a flat
homogeneous and isotropic Friedmann-Robertson-Walker (FRW)     metric
of the form 
\begin{equation}
ds^2=-dt^2+a^2(t)\left(dr^2+r^2d\Omega^2 \right),
\label{metric}
\end{equation}
 where $d\Omega^2\equiv d\theta^2+\sin^2\theta d\varphi^2$   is the solid angle,
  and $a(t)$ the scale factor. 
  In this setup, 
the 
first law of thermodynamics is interpreted in terms of the heat/energy that flows through the  apparent horizon of the Universe 
\citep{Jacobson:1995ab,Padmanabhan:2003gd,Padmanabhan:2009vy}, which in the case 
of flat geometry is just   
\citep{Bak:1999hd,Frolov:2002va,Cai:2005ra,Cai:2008gw}:
\begin{equation}
\label{apphor}
r_a=\frac{1}{H },
\end{equation}
  where $H=\dot a/a$ is the Hubble parameter (dots denote derivatives 
with respect to $t$).
Concerning  the  horizon temperature, 
this is given by the standard 
relation    \citep{Gibbons:1977mu}:
\begin{equation}
\label{Th}
 T=\frac{1}{2\pi r_a}.
\end{equation} 
Hence, the 
first law of thermodynamics is just
 $-dE=TdS$, where $-dE=A(\rho_m+p_m)H r_{a}dt$ is
the energy flow through the horizon during a time 
interval $dt$ in the case of a Universe filled with a matter perfect  fluid 
with energy density $\rho_m$ and pressure $p_m$  \citep{Cai:2005ra}.
  Differentiating  (\ref{kentropy}) we    obtain
\begin{equation}
\label{dsk}
dS_{K}=\frac{8\pi}{4G}\cosh{\left(K  \frac{\pi  }{GH^2}  
\right)}r_{a}\dot{r}_{a}dt,
\end{equation}
where we have used that $A=4\pi r_{a}^2=4 \pi/H^2 $.

Inserting   (\ref{Horentropy}), (\ref{Th}),  and 
(\ref{dsk}) into the first 
law of thermodynamics,  alongside with the relation $\dot r_{a}=-\dot{H}/H^2$,
  we  acquire   \citep{Lymperis:2021qty}
  \begin{equation} \label{gfe1}
-4\pi G(\rho_{m}+p_{m})=\dot{H}   \cosh{\left[K  
\frac{\pi}{GH^2}\right]}.
\end{equation}
Thus, using the matter conservation equation 
  \begin{equation}
 \dot{\rho}_m +3H(\rho_m +p_m)=0,
 \label{matterconsv}
 \end{equation}
the integration of (\ref{gfe1}) leads to    
\begin{eqnarray} \label{gfe2}
&&
\!\!\!\!\!\!\!\!\!\!\!\!\!\!\!\!
\frac{8\pi G}{3}\rho_{m}= H^{2} \cosh{\left[K  
\frac{\pi}{GH^2 }\right]} 
-\frac{K\pi}{G} \text{shi}{\left[K  
\frac{\pi}{GH^2 }\right]}-\frac{\Lambda}{3},
\end{eqnarray}
 with $\Lambda$   the integration constant and   ${\rm 
shi}(x)\equiv\int_0^{x}\sinh(x')/x'dx'$, a mathematical odd function of $x$ with 
no   discontinuities.

The above equations (\ref{gfe1}) and (\ref{gfe2}) are  the modified Friedmann 
equations in the scenario of Kaniadakis horizon entropy   cosmology  
\citep{Lymperis:2021qty}. 
As expected,  in the limit  
$K\rightarrow 0$ they turn into the standard ones. We can re-write them as   
\begin{align}
H^2 & =\frac{8\pi G}{3}(\rho_m+\rho_{DE}), \label{H1}
\\
    \dot{H}& =-4\pi G(\rho_m+\rho_{DE}+p_m+p_{DE}), \label{H2}
\end{align}
where we have introduced an effective dark-energy sector, with energy density 
and pressure respectively of the form 
\begin{small}
\begin{align}
&\rho_{DE}=\frac{3}{8\pi G}\Bigg\lbrace\frac{\Lambda}{3}+H^2\left[1-\cosh\left(\frac{\pi K}{GH^2}\right)\right]   +\frac{\pi K}{G}{\rm shi}\left(\frac{ \pi K}{GH^2}\right)\Bigg\rbrace, \\
&p_{DE}=-\frac{1}{8\pi 
G}\Bigg\lbrace\Lambda+(3H^2+2\dot{H})\left[1-\cosh\left(\frac{\pi 
K}{GH^2}\right)\right]    +\frac{3\pi K}{G}{\rm shi}\left(\frac{\pi 
K}{GH^2}\right)\Bigg\rbrace \label{PDE}. 
\end{align}
\end{small}
Therefore,  the EoS parameter for the dark energy 
sector  is
\begin{align}
 w_{DE}= & -1-2\dot{H}\left[1\!-\!\cosh\left(\frac{\pi 
K}{GH^2}\right)\!\right] \times \nonumber \\
& \left\lbrace\Lambda+3H^2\left[1\!-\!\cosh\left(\frac{
\pi K}{GH^2}\right)\! +\!\frac{3\pi K}{G}{\rm shi}\left(\frac{\pi 
K}{GH^2}\right)\right]\right\rbrace^{-1}.
\end{align}
In the general case, the cosmological equations are the two Friedmann equations 
 \eqref{H1} and \eqref{H2}, alongside the matter conservation equation 
(\ref{matterconsv}). \textbf{For convenience we focus on on the dust case, 
namely we consider  $p_m=0$}. It proves convenient to express the 
equations in terms of 
dimensionless variables. 
Introducing  the density parameters 
  \begin{align}
 \Omega_{\Lambda}\equiv \frac{\Lambda}{3 H^2}, 
\quad  \Omega_{m}\equiv \frac{8 \pi G \rho_m}{3 H^2},
  \end{align}
  the normalized Hubble function 
  \begin{align}
 E\equiv \frac{H}{H_0}, 
  \end{align}
  with $H_0$ the Hubble parameter at the present scale factor $a_0$, and
defining the
dimensionless  parameter $\beta$ as 
  \begin{align}
  \beta\equiv \frac{K \pi }{G H_0^2},
  \end{align}
  then, the cosmological equations are expressed as  
  \begin{align}
  &   \Omega_\Lambda^{\prime}(N)=  3   \Omega_\Lambda  \Omega_m
  \text{sech}\left(\frac{\beta }{E^2}\right),\\
  &   \Omega_m^{\prime}(N)=  3  
   \Omega_m \left[\Omega_m \text{sech}\left(\frac{\beta
   }{E^2}\right)-1\right], \\&   E^{\prime}(N)= -\frac{3}{2} E  
   \Omega_m \text{sech}\left(\frac{\beta
   }{E^2}\right),
  \end{align}
  where primes denote derivatives with respect to the e-foldings
  number $N= \ln (a/a_0)$  (and thus $    f^{\prime}  = 
 \dot{f}/H$). Note that using the above variables, the first Friedmann equation 
 \eqref{H1} gives rise to  the constraint
 \begin{align}
\beta  \text{shi}\left(\frac{\beta }{E^2}\right)+E^2
   \left[-\cosh \left(\frac{\beta }{E^2}\right)+\Omega_\Lambda
   + \Omega_{m}\right]=0,\label{eqN4.9}
 \end{align} 
 which allows us to eliminate $\Omega_\Lambda$ in terms of $\Omega_{m}$ and $E$.
 Finally, note that for the effective dark energy density parameter, in the 
general case we have
\begin{equation}
    \Omega_{DE}= 1-\Omega_m = 1 + \beta  \text{shi}\left( \beta  
\right) -\cosh \left( \beta  \right)+ \Omega_\Lambda . 
\label{constreq1}
\end{equation}
Lastly, it proves convenient to introduce the 
  deceleration parameter $q(z)$, and   the cosmographic 
jerk parameter $j(z)$, which are defined as
\begin{align}
q :=& -1- \frac{E^{\prime}}{E}, \label{q}\\
j:= & q(2q+1)-q', \label{j}
\end{align}
where $j=1$ recovers the case of a  cosmological constant.

In the scenario of   Kaniadakis horizon entropy   cosmology  one may have 
the general integration constant $\Lambda$, which will play the role of an 
explicit cosmological constant, or one may set it to zero and thus require for 
the extra $K$-dependent terms to drive the Universe acceleration. Since the 
corresponding equation structure (which will be  used later for the  the 
Bayesian statistical analysis and the  dynamical 
system approach)  is  different in the two cases, we 
examine them separately in the following subsections.

\subsection{Case I: $\Lambda \neq 0$} 

In the general case where $\Lambda \neq 0$, i.e.  $\Omega_\Lambda\neq 0$, we 
  use  the constraint equation (\ref{eqN4.9})  to obtain the reduced dynamical 
system
\begin{align}
      &   \Omega_m^{\prime}(N)=  3  
   \Omega_m \left[\Omega_m \text{sech}\left(\frac{\beta
   }{E^2}\right)-1\right], \label{evolm}\\&   E^{\prime}(N)= -\frac{3}{2} E  
   \Omega_m \text{sech}\left(\frac{\beta
   }{E^2}\right). \label{evolE}
\end{align}
This is integrable, with 
\begin{small}
\begin{align}
 & \Omega_{m}(E)= \cosh \left(\frac{\beta
   }{E^2}\right)+\frac{-\cosh (\beta )-\beta 
   \text{shi}\left(\frac{\beta }{E^2}\right)+\beta 
   \text{shi}(\beta )+ \Omega_m^{(0)}}{E^2}, \\
   & \Omega_m(1)=\Omega_{m}^{(0)},
\end{align}
\end{small}
and 
\begin{small}
\begin{align}
    & E^{\prime}(N) =  -\frac{3}{2} E   -\frac{3   \text{sech}\left(\frac{\beta }{E^2}\right)
   \left(-\cosh (\beta )-\beta  \text{shi}\left(\frac{\beta
   }{E^2}\right)+\beta  \text{shi}(\beta )+ \Omega_m^{(0)}\right)}{2 E}, \label{4.13}
\end{align}
\end{small}
where $\Omega_m(N=0)=\Omega_{m}^{(0)}$ and  $E(N=0)=1$. 

The equation 
\eqref{4.13} is easily integrated to give 
\begin{small}
\begin{align}
  {3}  N (E)= &  -\ln
   \Bigg( 1-\frac{\cosh (\beta )-E^2  \cosh \left(\frac{\beta
   }{E^2}\right)+\beta  \text{shi}\left(\frac{\beta
   }{E^2}\right)-\beta  \text{shi}(\beta )}{\Omega_m^{(0)}}\Bigg),
\end{align}
\end{small}
introducing as dynamical variable the redshift 
  $ e^N= a= (1+z)^{-1}$, where $z=0$ and $a_0=1$ for current time, the previous equation leads to 
  \begin{small}
 \begin{align}
     &  (z+1)^{{3 }}=   1-\frac{\left[\cosh (\beta )-E^2  \cosh \left(\frac{\beta
   }{E^2}\right)+\beta  \text{shi}\left(\frac{\beta
   }{E^2}\right)-\beta  \text{shi}(\beta )\right]}{\Omega_m^{(0)}}. \label{eqN4.17}
\end{align}
\end{small}
Evaluating \eqref{constreq1} at present time  
 gives
\begin{equation}
    \Omega_{DE}^{(0)}= 1-\Omega_m^{(0)}= 1 + \beta  \text{shi}\left( \beta  
\right) -\cosh \left( \beta  \right)+ \Omega_\Lambda^{(0)}, 
\label{exact3.20aa}
\end{equation}
and combining it with \eqref{eqN4.17} we have 
\begin{align}
     & \Omega_m^{(0)}  (z+1)^{3}+ \Omega_\Lambda^{(0)}=E^2  \cosh \left(\frac{\beta
   }{E^2}\right) - \beta  \text{shi}\left(\frac{\beta
   }{E^2}\right). \label{exact3.20}
\end{align}

We mention here that in the general case where $\Lambda\neq 0$ from the above 
we obtain the relation between $\beta$, $ \Omega_m^{(0)}$  and $ 
\Omega_\Lambda^{(0)}$ as 
\begin{equation}
\label{exact3.20aan}
\Omega_m^{(0)}= \cosh \left( \beta  \right) - \beta  \text{shi}\left( \beta  
\right) - \Omega_\Lambda^{(0)}.
\end{equation}

 We expand (\ref{exact3.20aa}), (\ref{exact3.20})  
  up to third order around $\beta=0$, resulting in 
\begin{equation}
  \Omega_m^{(0)}+\Omega_\Lambda^{(0)}=1-\frac{\beta ^2}{2}+O\left(\beta 
^4\right),
\end{equation}
and 
\begin{align}
 \Omega_m^{(0)}  (z+1)^{{3 }}+ \Omega_\Lambda^{(0)}& \approx E^2-\frac{\beta 
^2}{2 E^2}+O\left(\beta ^4\right).
\end{align}
Hence, we obtain four roots:
\begin{small}
\begin{align}
&E_{1,2}= \mp\frac{\sqrt{\Omega_m^{(0)}  (z+1)^{{3 }}+ 
\Omega_\Lambda^{(0)}-\sqrt{2 \beta ^2+\left[\Omega_m^{(0)}  (z+1)^{{3 }}+ 
\Omega_\Lambda^{(0)}\right]^{2}}}}{\sqrt{2}}, \\
&E_{3,4}=\mp\frac{\sqrt{\Omega_m^{(0)}  (z+1)^{{3 }}+ 
\Omega_\Lambda^{(0)}+\sqrt{2 \beta ^2+\left[\Omega_m^{(0)}  (z+1)^{{3 }}+ 
\Omega_\Lambda^{(0)}\right]^{2}}}}{\sqrt{2}}. \label{E4}
\end{align}
\end{small}
Solutions $E_1$ and $E_2$ are complex,  and $E_3$ is negative. Thus, 
the only physical solution is $E_4$. 
In the following, instead of using the exact implicit formula for $E$ 
given in \eqref{exact3.20}, we will consider the approximation $E_4$ in 
\eqref{E4}.

\subsection{Case II: $\Lambda = 0$} 

In the case where an explicit cosmological constant is absent, namely $\Lambda 
= 0$, i.e  $\Omega_\Lambda=0$, the general system  \eqref{evolm}, \eqref{evolE}
reduces to  
\begin{equation}
    E^{\prime}(N)=-\frac{3}{2} E   + \frac{3 \beta    \text{sech}\left(\frac{\beta
   }{E^2}\right) \text{shi}\left(\frac{\beta
   }{E^2}\right)}{2 E}.
\end{equation}
The last equation is easily integrated to give 
\begin{align}
   N (E)= -\frac{1}{3}\ln \left[\frac{E^2 \cosh \left(\frac{\beta }{E^2}\right)-\beta  \text{shi}\left(\frac{\beta }{E^2}\right)}{\cosh (\beta )-\beta  \text{shi}(\beta )}\right],
\end{align}
which, using the redshift, 
implies
 \begin{align}
     &    (z+1)^{{3 }}=\frac{E^2 \cosh \left(\frac{\beta }{E^2}\right)-\beta  
\text{shi}\left(\frac{\beta } {E^2}\right)}{\cosh (\beta )-\beta  
\text{shi}(\beta )}. \label{eqN20}
\end{align}
Hence, using \begin{equation}
\label{corr}
     \Omega_m^{(0)}= {\cosh \left( \beta \right)-\beta  \text{shi}\left(\beta\right)},
\end{equation} 
as it arises from \eqref{exact3.20aan} for $\Omega_\Lambda^{(0)}=0$, 
 we  obtain
  \begin{align}
     &   \Omega_m^{(0)} (z+1)^{{3 }}= {E^2 \cosh \left(\frac{\beta }{E^2}\right)-\beta  \text{shi}\left(\frac{\beta }{E^2}\right)}. \label{eqN3.34}
\end{align}
 Expanding (\ref{eqN3.34})  
  up to third order around $\beta=0$, we result to 
\begin{align}
 \Omega_m^{(0)} (z+1)^{{3 }} & \approx E^2-\frac{\beta ^2}{2 
E^2}+{\mathcal{O}}\left(\beta ^4\right),
 \end{align}
 and thus at present times gives $
  \Omega_m^{(0)}=  1-\frac{\beta ^2}{2}+{\mathcal{O}}\left(\beta ^4\right)$.
 Therefore, we obtain four roots:
 \begin{small}
\begin{eqnarray}
&& E_{1,2}= \mp\frac{\sqrt{{\Omega_m^{(0)}} (z+1)^{ 3}-\sqrt{2 \beta 
^2+{\Omega_m^{(0)}}^2 (z+1)^{6}}}}{\sqrt{2}},\\
&&E_{3,4}=\mp\frac{\sqrt{{\Omega_m^{(0)}} (z+1)^{ 3}+\sqrt{2 \beta 
^2+{\Omega_m^{(0)}}^2 (z+1)^{6}}}}{\sqrt{2}}. \label{eqN340}
\end{eqnarray}
\end{small}
Similarly to the previous case, roots $E_1$ and $E_2$ are complex   
 while $E_3$ is negative. 
Consequently, the only physical solution is $E_4$ in \eqref{eqN340}.

\section{Observational constraints} \label{sec:data}

In this section we confront the scenario of Kaniadakis horizon-entropy 
cosmology with observations. We are interested in extracting the bounds on the 
parameter phase-space $\Theta = \{h, \Omega_m^{(0)}, \beta\}$ and $\{h, 
\beta\}$, 
particularly on the  parameter $\beta$, which is related to the 
Kaniadakis basic parameter $K$.  For 
convenience, we focus on the physically interested case of  dust matter, namely 
we set $w_m=0$.

\subsection{Datasets and methodology} \label{subsec:data}
 
We will employ the most commonly used datasets.

\begin{itemize}
    \item {\it  Observational Hubble Data} (OHD).  
    The sample contains $31$ cosmological-independent measurements of the Hubble parameter in the redshift range $0.07<z<1.965$ from passive elliptic galaxies,  the so-called cosmic chronometers \citep{Moresco:2016mzx}.
    
    \item {\it Pantheon Supernova Type Ia sample} (SNIa). We use 1048 
data points of the distance modulus, $\mu(z)_{\mathrm{SNIa}}$, of high-redshift 
SNIa  in the redshift range $0.001<z<2.3$ \citep{Scolnic:2018}.
    
    \item {\it HII galaxies} (HIIG).
    It contains a total of $181$ data points of the distance modulus $\mu_{\mathrm{HIIG}}(z)$ estimated from the Balmer line luminosity-velocity dispersion relation for HII galaxies spanning the redshift region $0.01<z<2.6$ \citep{Gonzalez-Moran:2021drc}.
    
    \item {\it Strong lensing systems} (SLS). We use the sample by \citet{Amante_2020} which contains $143$ strong lensing systems by elliptical galaxies with measurements of the redshift for the lens and the source, spectroscopic velocity dispersion and the Einstein radius. 
    These quantities 
    allow us to construct an observational distance ratio within the region $ 0.5 \leq D^{obs} \leq 1$ .
    
    \item {\it Baryon acoustic oscillations} (BAO). We consider $6$ correlated data points of the imprint of baryon acoustic oscillation in the size of the sound horizon in clustering and power spectrum of galaxies measured by \citet{Percival:2010,Blake:2011,Beutler:2011hx} and collected by \citet{Giostri:2012}.
\end{itemize}

We would like to mention here  that other 
cosmological observations could be included in the parameter estimation too, 
for instance the CMB data. To perform such analysis in a robust 
way, a full perturbation approach is needed in order to obtain the linear 
Einstein-Boltzmann equations. Nevertheless this is  beyond the scope of the 
present  work.
An alternative approach would be  to use the distance priors from Planck 2018 
based on slight deviations from $\Lambda$CDM, such as the $w$CDM model 
\citep{Chen:2019JCAP}.  However, since this procedure could lead to biased 
constraints, in the following we prefer not to use the CMB dataset.

The inference of the cosmological parameters under Kaniadakis  horizon entropy 
 cosmology for both scenarios  ($\Lambda\neq0$ and $\Lambda=0$)
is performed by   a Bayesian Markov Chain Monte Carlo (MCMC) approach and the \texttt{emcee} Python module \citep{Emcee:2013}. We set $3000$ chains with $250$ steps each,   and consider uniform priors in the ranges: $h:[0.2, 1]$, $\Omega_m^{(0)}:[0,1]$, $\beta: [-\pi, \pi]$. The burn-in phase is stopped up to obtain convergence according to the auto-correlation time criterion. 
Then, we build a Gaussian  
log-likelihood as the merit-of-function to minimize 
through the equation
$-2\ln(\mathcal{L})\varpropto \chi^2$, where $\chi$ is the chi-square function given by
\begin{equation}
    \chi^2_{\rm uncorr} = 
\sum_{i}^{N_{dat}}\left(\frac{\mathcal{D}-\mathcal{M}}{\sigma_{\mathcal{D}}}
\right)^2\,,
\end{equation}
for the samples OHD, HIIG, SLS because the measurements are considered to be uncorrelated. $N_{dat}$ is the number of points of dataset $\mathcal{D}$, $\sigma_{\mathcal{D}}$ is the estimated uncertainty  for each dataset, and $\mathcal{M}$ represents the theoretical quantity of that observable \textbf{based on $E_4$ presented in \eqref{E4} and \eqref{eqN340} for $\Lambda\neq 0$ and $\Lambda=0$ models respectively.} As SNIa and BAO datasets contain correlated points, the figure of merit is built as
\begin{equation}
    \chi^{2}_{\rm corr} = \Delta\vec{x}\cdot{\rm C}^{-1}\cdot \Delta\vec{x}^{T}\,,
\end{equation}
where $\Delta\vec{x}$ is the difference between the observational and theoretical quantities, and $\rm C^{-1}$ is the covariance matrix. It is worth to mention that a nuisance parameter is presented in the SNIa data and it is convenient to marginalize over it to reduce the uncertainties. Thus, the figure of merit for SNIa data is  
\begin{equation}
    \chi^2_{\rm SNIa} = a + \ln\left( \frac{e}{2\pi}\right) - \frac{b^2}{e}\,,
\end{equation}
where $a$, $b$, and $e$ are functions of $\Delta\vec{x}$ and $\rm C^{-1}$. For 
more details on these expressions see \citet{Motta:2021hvl}.

Finally, we perform a joint analysis through the sum of the function-of-merits 
of each data sample, namely
\begin{equation}
    \chi^2_{\rm Joint}=\chi^2_{\rm OHD}+\chi^2_{\rm SLS}+\chi^2_{\rm HIIG}+\chi^2_{\rm SNIa}+\chi^2_{\rm BAO},
\end{equation}
where subscripts indicate the dataset under consideration.

\subsection{Results  } 

\begin{figure*}
    \centering
    \includegraphics[width=0.6\textwidth]{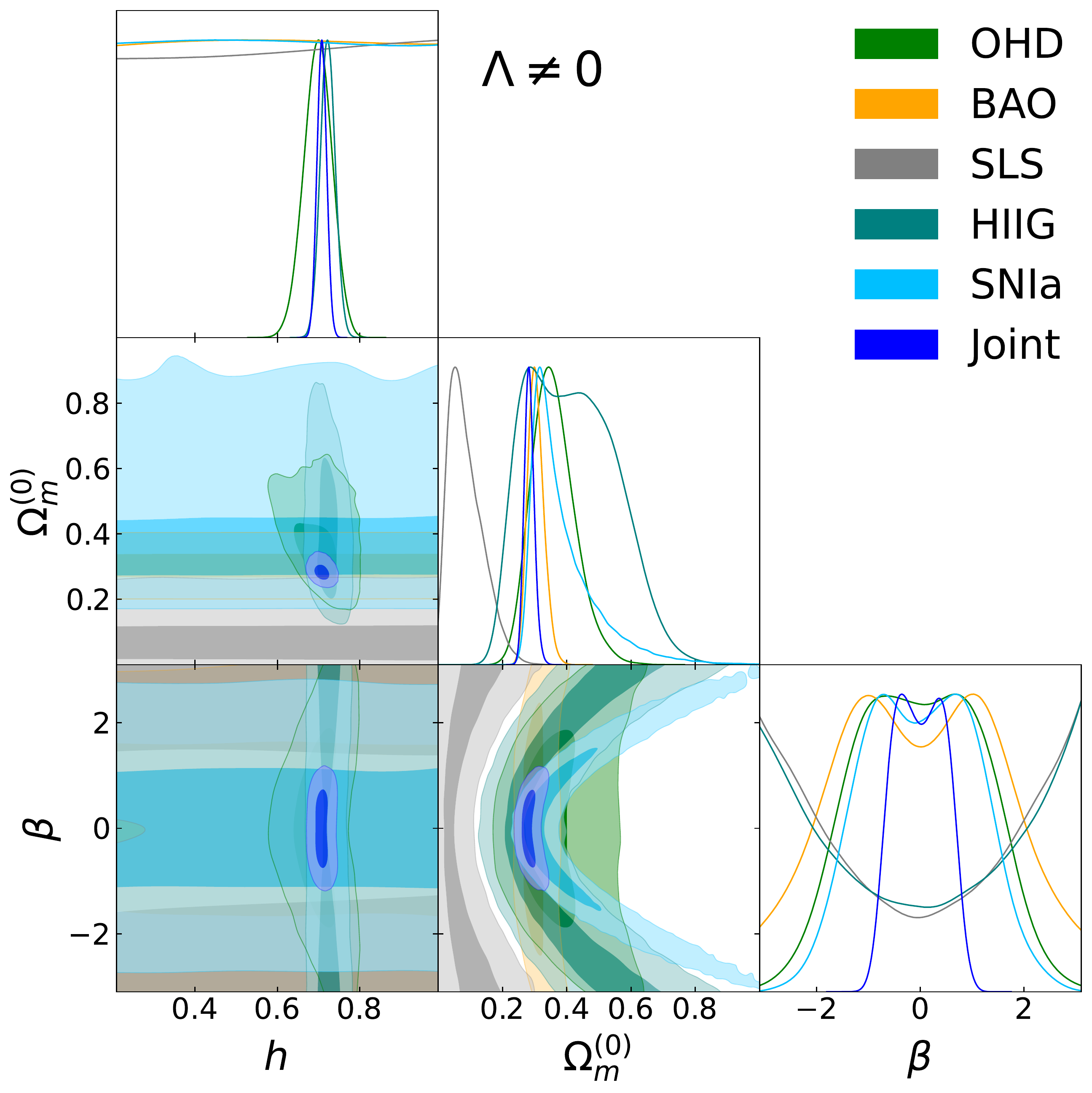}\\
    \includegraphics[width=0.6\textwidth]{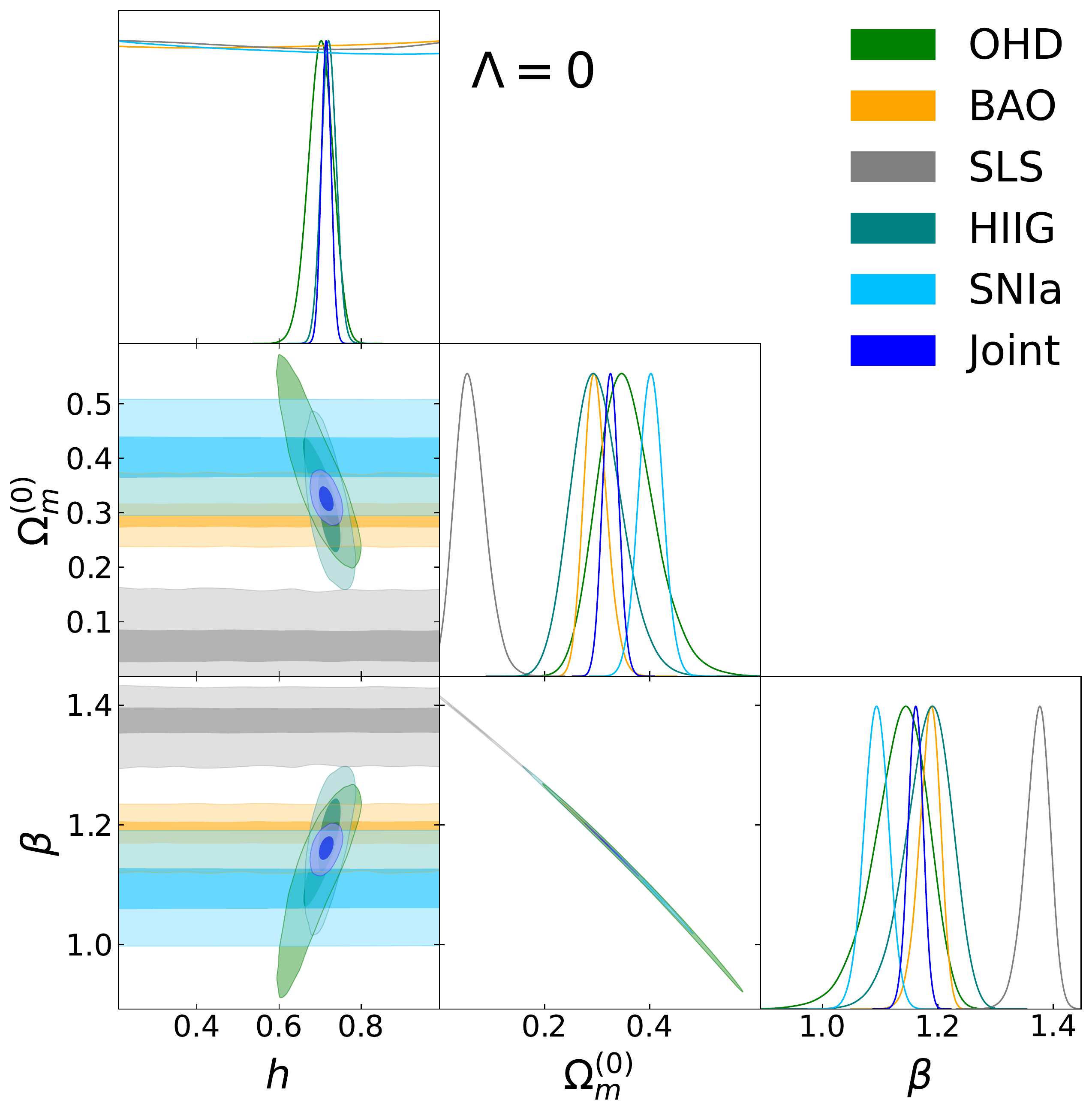}
    \caption{
      Two-dimensional log-likelihood contours at 68\% and 99.7\%  confidence level  
(CL), alongside the corresponding 1D posterior distribution of the free 
parameters, in Kaniadakis horizon entropy cosmology, for   $\Lambda\neq0$ 
(upper panel) and $\Lambda=0$ 
(lower panel). We use the various 
datasets described in the text, as well as the joint analysis.
 }
    \label{fig:contours}
\end{figure*}

Performing the full confrontation of the scenario  we construct the 
corresponding log-likelihood contours  at $68\%$ ($1\sigma$) and 
$99.7\%$ ($3\sigma$) confidence level (CL), and we present them in 
Fig. \ref{fig:contours} alongside the 1D posterior 
distribution.
Moreover, in Table \ref{tab:bestfits} we 
show the mean values and the 
uncertainties at $1\sigma$   confidence level for the parameters $h$, 
$\Omega_m^{(0)}$ and $\beta$ for both  $\Lambda \neq 0$ 
and $\Lambda=0$ cases.   

\begin{table*}
\centering
\begin{tabular}{|lccccccc|}
\hline
Sample &    $\chi^2_{\rm min}$     &  $h$ & $\Omega_m^{(0)}$ & $\beta$ & $\Delta$AICc 
& $\Delta$BIC & $\Delta$DIC \\
\hline
\multicolumn{8}{|c|}{Case $\Lambda \neq 0$} \\ [0.9ex]
OHD    & $19.25$   & $0.699^{+0.033}_{-0.034}$  & $0.354^{+0.072}_{-0.061}$  & 
$-0.004^{+1.259}_{-1.255}$ & $7.6$ & $8.2$   & $-4.5$\\ [0.9ex] 
BAO  & $2.91$    & $0.599^{+0.272}_{-0.270}$  & $0.302^{+0.027}_{-0.023}$  & 
$-0.016^{+1.596}_{-1.594}$ & $14.1$ & $1.9$  & $0.2$\\ [0.9ex] 
SLS   & $216.52$  & $0.608^{+0.268}_{-0.277}$  & $0.077^{+0.064}_{-0.042}$  & 
$-0.006^{+2.550}_{-2.526}$ & $5.5$ & $8.3$  & $-5.2$ \\ [0.9ex] 
HIIG  & $452.96$  & $0.722^{+0.018}_{-0.018}$  & $0.408^{+0.151}_{-0.137}$  & 
$0.043^{+2.510}_{-2.562}$  & $19.3$ & $22.3$ & $-19.6$\\ [0.9ex] 
SNIa  & $1042.99$ & $0.598^{+0.273}_{-0.270}$  & $0.359^{+0.126}_{-0.055}$  & 
$0.009^{+1.108}_{-1.124}$  & $9.0$ & $14.0$  & $-14.6$\\ [0.9ex] 
Joint & $1743.48$ & $0.708^{+0.012}_{-0.011}$  & $0.283^{+0.016}_{-0.015}$  & 
$-0.011^{+0.517}_{-0.507}$  & $2.9$ & $8.1$  & $0.5$ \\ [0.9ex] 
\hline
\multicolumn{8}{|c|}{Case $\Lambda = 0$} \\ 
OHD    & $14.56$   & $0.701^{+0.029}_{-0.030}$  & $0.353^{+0.057}_{-0.050}$  & 
$1.138^{+0.043}_{-0.051}$  & $0.5$ & $0.0$  & $0.0 $ \\ [0.9ex] 
BAO    & $2.33$    & $0.602^{+0.272}_{-0.273}$  & $0.297^{+0.023}_{-0.021}$  & 
$1.186^{+0.017}_{-0.020}$  & $3.6$ & $-0.4$  & $-0.3 $ \\ [0.9ex] 
SLS    & $212.86$  & $0.596^{+0.276}_{-0.270}$  & $0.057^{+0.031}_{-0.027}$  & 
$1.373^{+0.020}_{-0.023}$  & $-0.2$ & $-0.3$ & $0.0 $\\ [0.9ex] 
HIIG   & $435.64$  & $0.721^{+0.018}_{-0.018}$  & $0.298^{+0.050}_{-0.045}$  & 
$1.185^{+0.038}_{-0.043}$  & $-0.1$ & $-0.2$ & $-0.1 $\\ [0.9ex] 
SNIa   & $1036.48$ & $0.596^{+0.275}_{-0.271}$  & $0.402^{+0.023}_{-0.023}$  & 
$1.093^{+0.021}_{-0.021}$  & $0.5$ & $0.5$   & $0.5 $ \\ [0.9ex] 
Joint  & $1753.03$ & $0.715^{+0.012}_{-0.012}$  & $0.326^{+0.015}_{-0.015}$  & 
$1.161^{+0.013}_{-0.013}$  & $10.4$ & $10.4$ & $10.4 $\\ [0.9ex] 
\hline
\end{tabular}
\caption{Best-fit values and their $68\%$ CL uncertainties for Kaniadakis 
horizon entropy cosmology with $\Lambda\neq0$ (upper panel) and $\Lambda=0$ 
(lower panel) employing the data sets: OHD ($31$ data points), BAO (6 data points), SLS (143 data points), HIIG (181 data points), SNIa (1048 data points) and the joint analysis of them.  }
\label{tab:bestfits}
\end{table*}

\begin{figure*}
\centering
\includegraphics[width=0.31\textwidth]{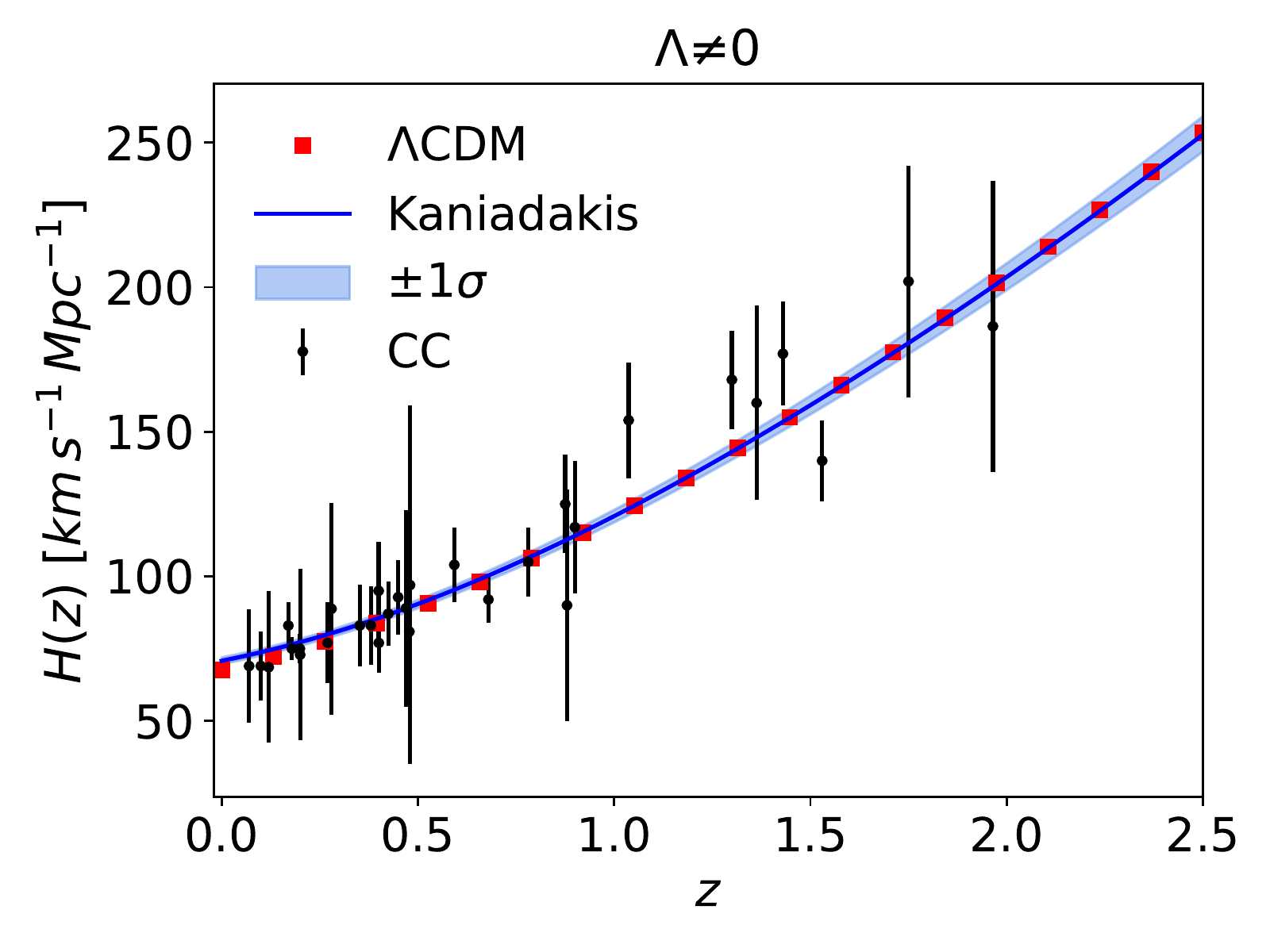}
\includegraphics[width=0.31\textwidth]{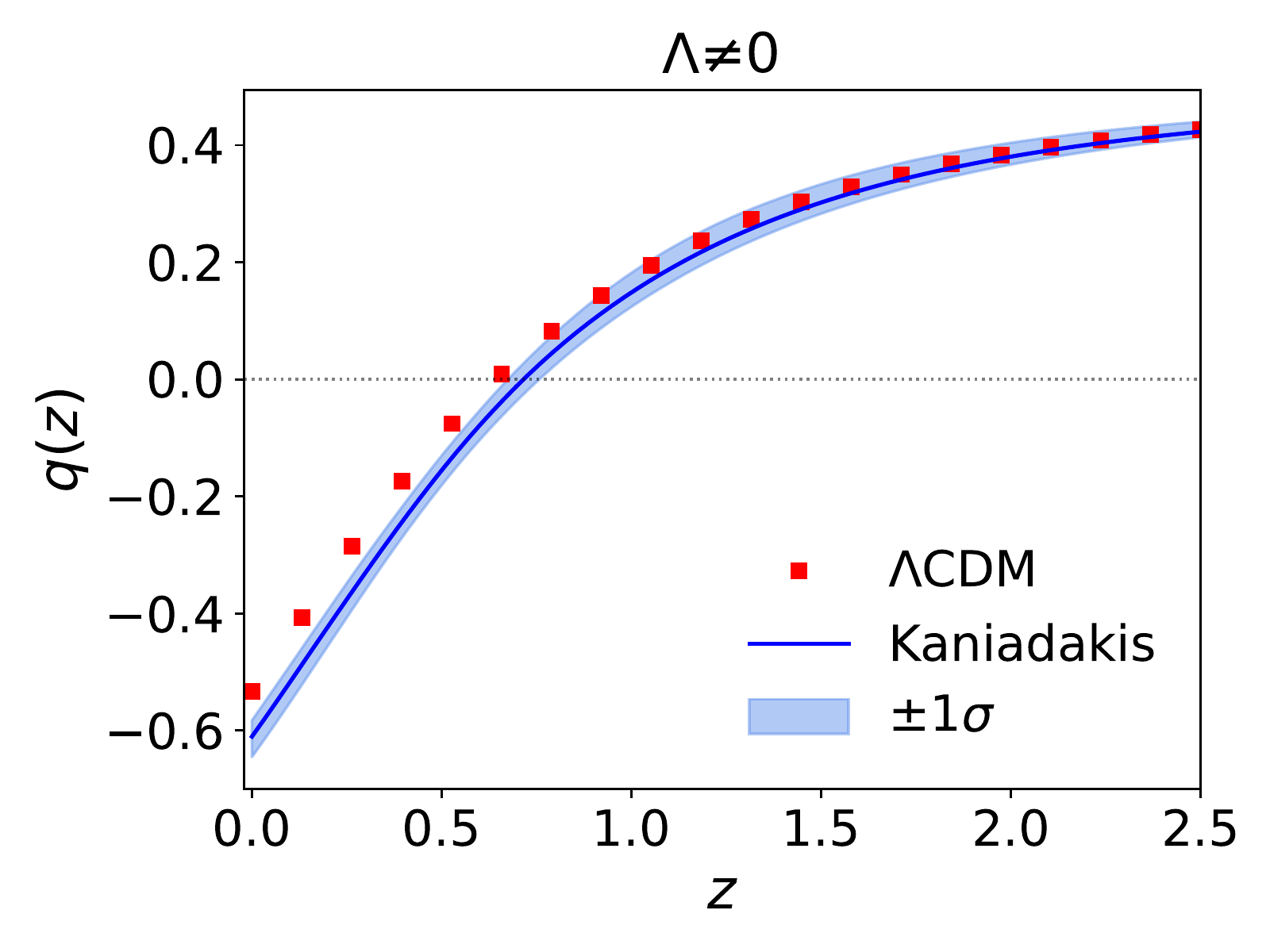}
\includegraphics[width=0.31\textwidth]{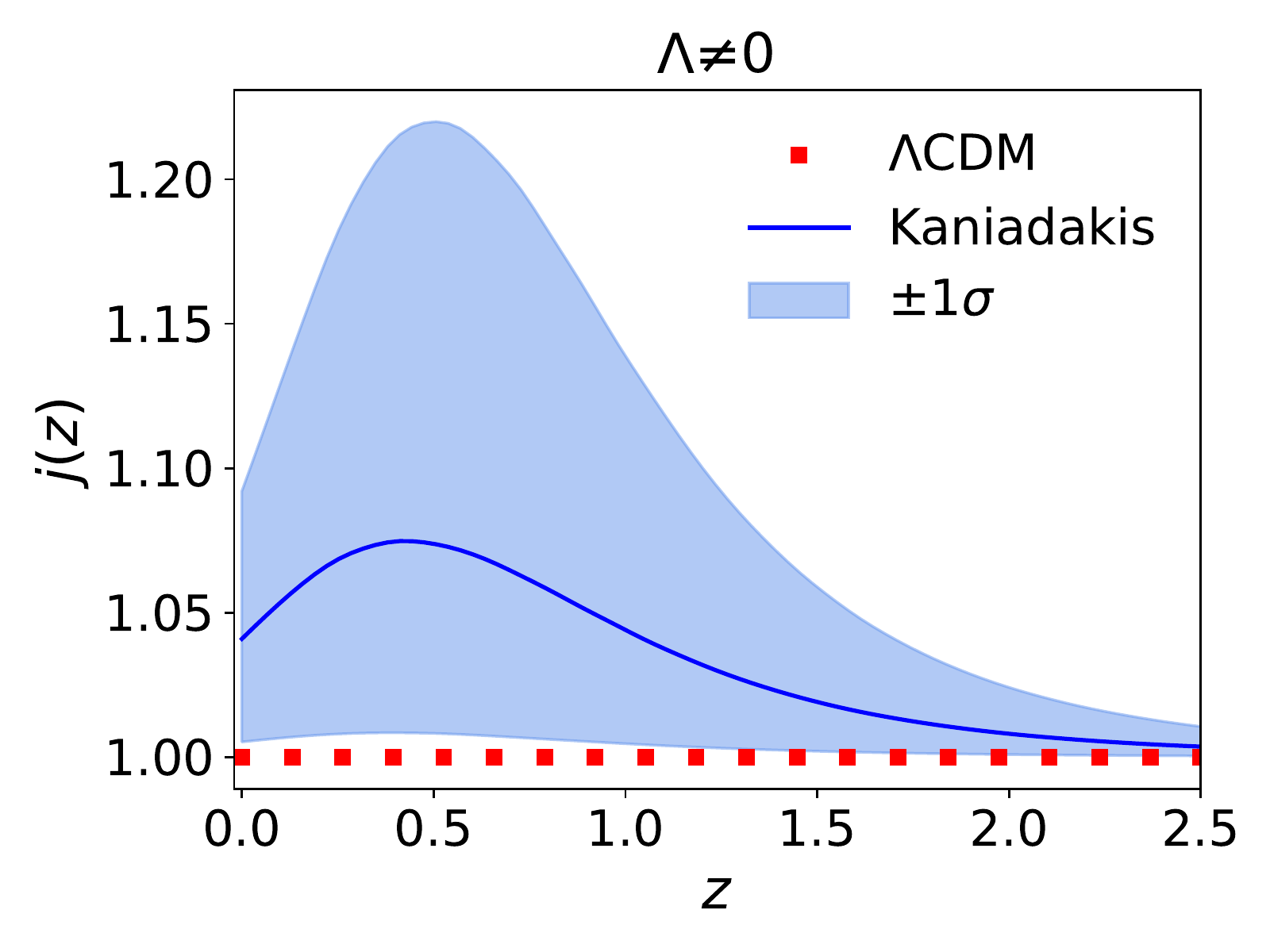}\\
\includegraphics[width=0.31\textwidth]{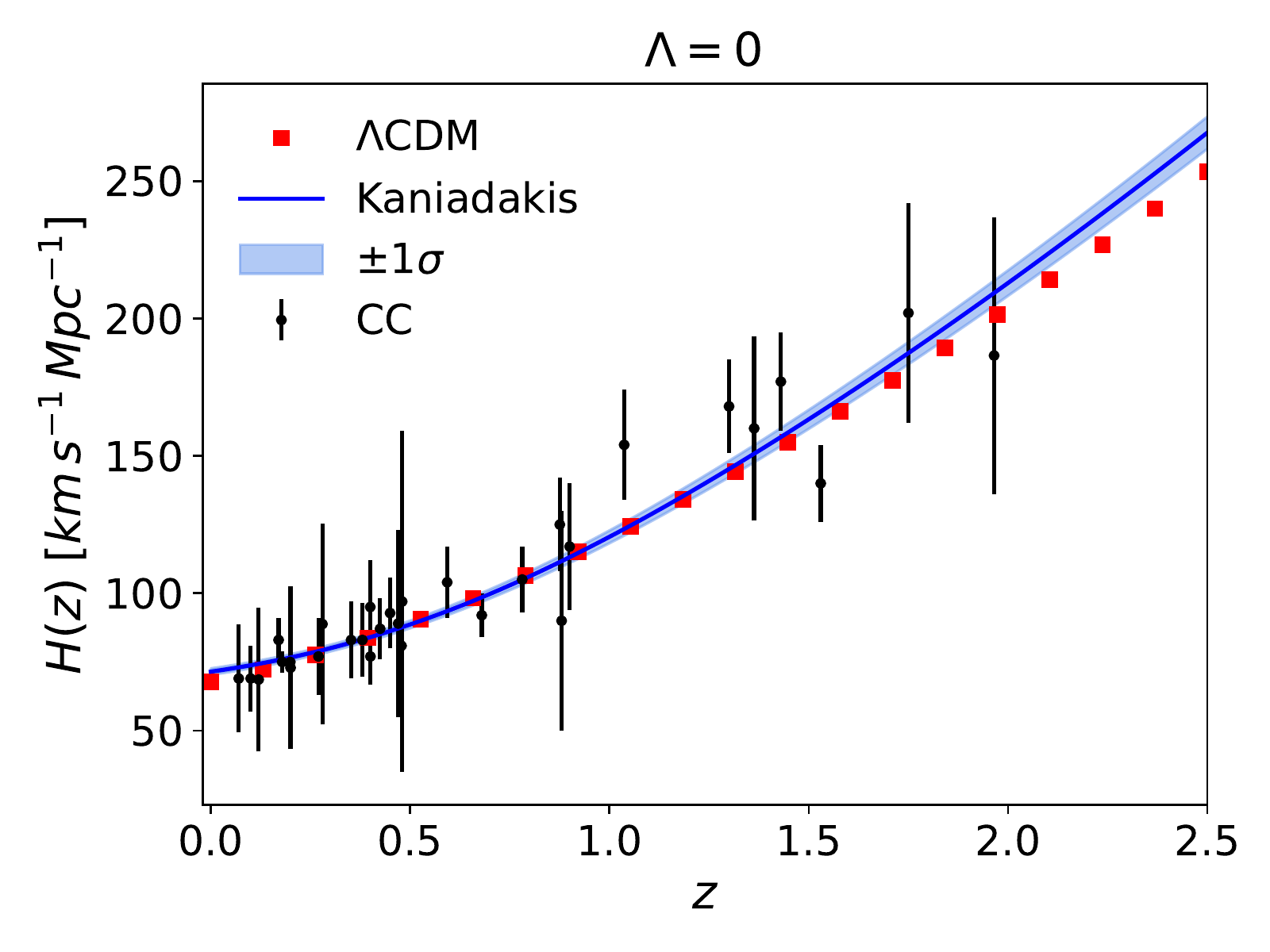}
\includegraphics[width=0.31\textwidth]{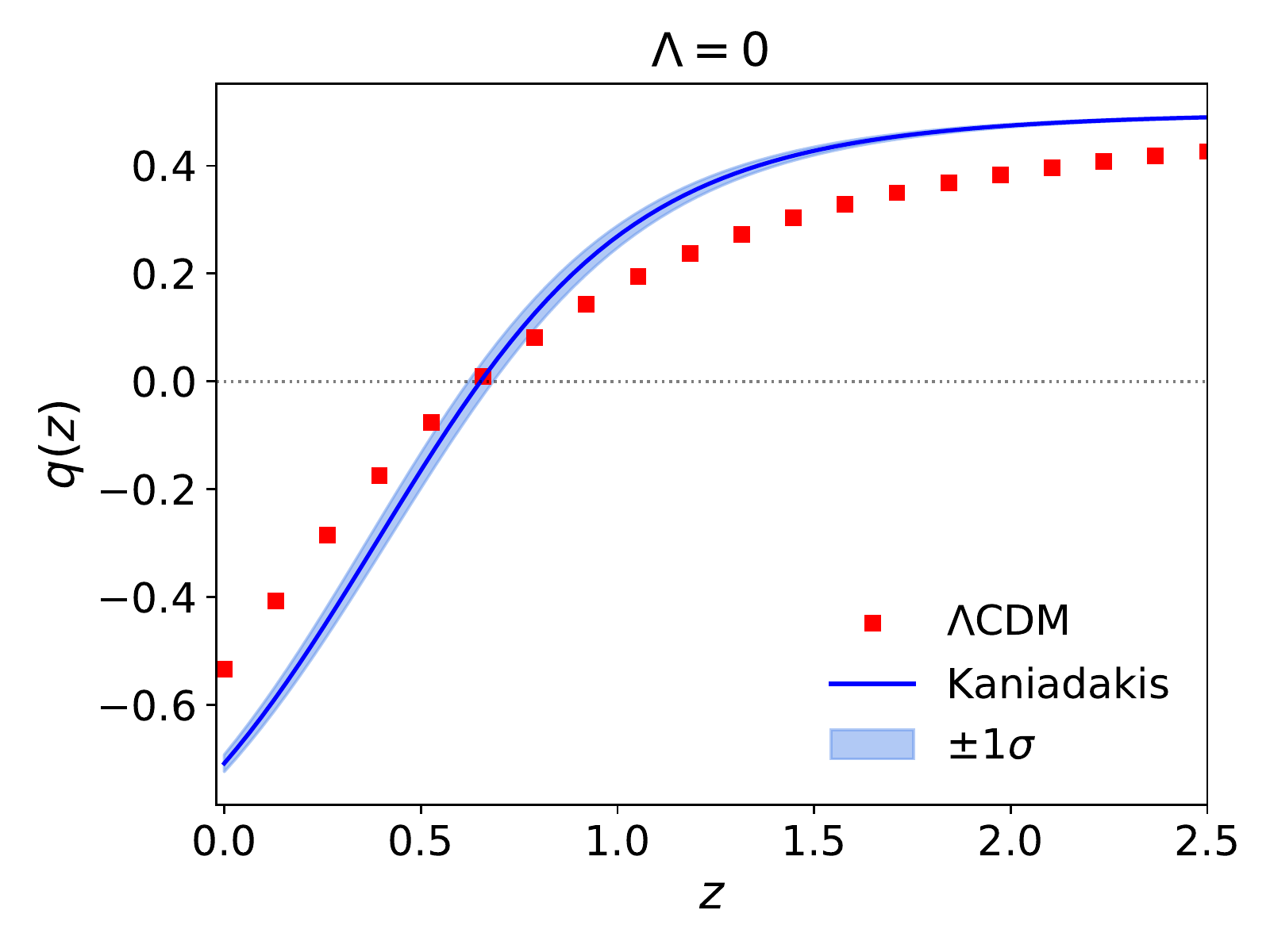}
\includegraphics[width=0.31\textwidth]{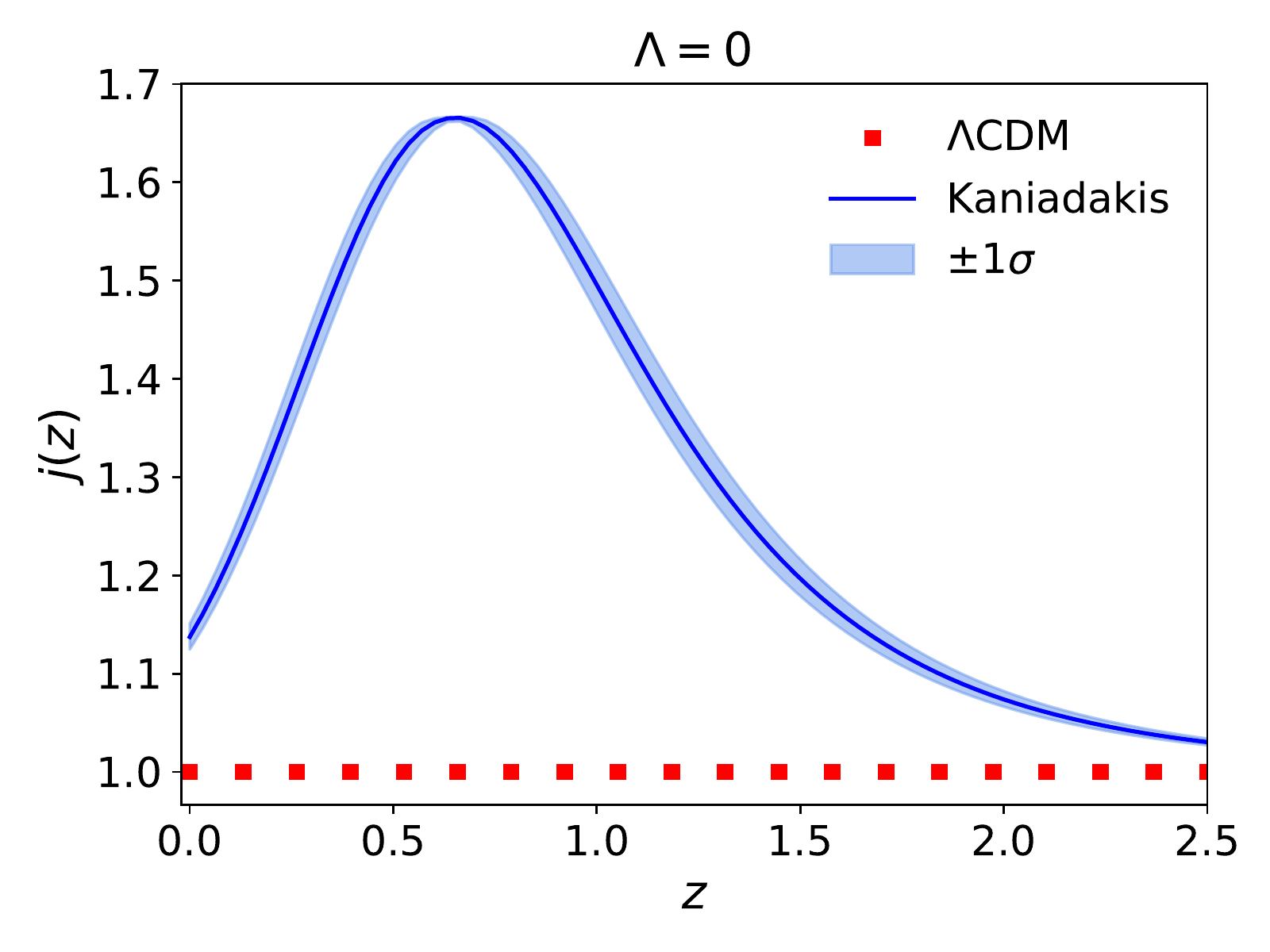}\\
\caption{Upper panel, left to right: reconstruction of the $H(z)$, 
$q(z)$, 
and $j(z)$, 
in Kaniadakis horizon entropy cosmology with  $\Lambda 
\neq 0$. Lower panel: 
same as before for the case $\Lambda = 0$. We have used the bound obtained from the joint 
analysis, and the shaded regions denote the uncertainties at $1\sigma$. 
For completeness, the red square-points 
represent the results of  $\Lambda$CDM cosmology  with $h=0.6766$ and
$\Omega_m^{(0)}=0.3111$ 
\citep{Planck:2018vyg}.}
\label{fig:Hz_and_qz}
\end{figure*}

As  
can be seen, the 
bounds estimated from each data sample are consistent among  
themselves,   
although the SLS dataset provides lower values for $\Omega_m^{(0)}$. The joint constraints 
$h=0.708^{+0.012}_{-0.011}$ ($h=0.715^{+0.012}_{-0.012}$) for the case $\Lambda 
\neq 0$ ($\Lambda=0$), are consistent at $2.67\sigma$ ($3\sigma$) with the one 
estimated from the CMB anisotropies \citep{Planck:2018vyg} and at  $1.74\sigma$ 
($1.36\sigma$) with the one from SH0ES \citep{Riess:2019}. 
Hence, the scenario of Kaniadakis horizon entropy cosmology can offer an 
alleviation to the $H_0$ tension, providing  a value in between its   local 
measurements, and it indirect estimation from the early stages of the Universe. 

Concerning Kaniadakis parameter, we
find that, when $\Lambda \neq 
0$, the combination of the data samples constrains 
$\beta=-0.011^{+0.515}_{-0.507}$, namely   $\beta$ is constrained 
around 0 as expected,  
the value in which Kaniadakis entropy 
becomes the standard Bekenstein-Hawking one.
However,  when $\Lambda=0$, the joint constraint yields $\beta=1.161^{+0.013}_{-0.013}$,  
which is expected, as we mentioned above,  
because in the absence of an explicit 
cosmological constant one needs a significant deviation from standard 
cosmology to describe the Universe acceleration.
Finally, note that due to equation \eqref{corr} that holds in the $\Lambda=0$ 
case, we acquire a
correlation between $ \Omega_m^{(0)}$ and the Kaniadakis parameter $\beta$ in 
the lower panel of Fig. \ref{fig:contours}.

Let us make a comment on the predicted entropy today, since this is 
possible to be calculated through Eq. \eqref{kentropy}. According to our 
model, and imposing for the horizon area of our Universe its present value, we 
arrive at the value $S_K\sim1.44\times10^{99}$ m$^2$Kg s$^{-2}$K$^{-1}$ 
(for $\Lambda\neq0$) and $S_K\sim3.15\times10^{99}$m$^2$Kg s$^{-2}$K$^{-1}$ 
(for $\Lambda=0$). In comparison, for the standard Bekenstein-Hawking  
entropy, we have $S_{BH}\sim2.83\times10^{99}$ m$^2$Kg s$^{-2}$K$^{-1}$ 
($\Lambda\neq0$) and $S_{BH}\sim2.79\times10^{99}$m$^2$Kg s$^{-2}$K$^{-1}$ 
($\Lambda=0$). Therefore, the corresponding ratio is $S_K/S_{BH}\simeq0.5$ for 
 $\Lambda\neq0$ and $S_K/S_{BH}\simeq1.12$ for $\Lambda=0$, which implies a 
small difference between Kaniadakis and Bekenstein-Hawking entropies.
 
Due to the competitive qualities of the fits obtained from both scenarios, it 
would be interesting to statistically compare them
with the concordance 
$\Lambda$CDM cosmology. In order to achieve this, we apply the standard 
  criteria, namely the Akaike information criterion corrected for small samples  
  \citep[AICc,][]{AIC:1974, 
Sugiura:1978, AICc:1989} and 
the Bayesian information criterion
\citep[BIC,][]{schwarz1978}, since 
$\Lambda\neq 0$ model contains one extra free parameter over $\Lambda$CDM. The 
AICc and BIC are defined as ${\rm AICc}= \chi^2_{\text{min}}+2k +(2k^2+2k)/(N-k-1)$ and 
${\rm BIC}=\chi^2_{\text{min}}+k\ln(N)$ respectively,   where $\chi^2_{\text{min}}$ is the 
minimum of the $\chi^2$ function, $N$ is the size of the dataset and $k$ is the 
number of free parameters. Following the rules described in 
\citet{hernandezalmada2021kaniadakis}, we find that $\Lambda=0$ model and 
$\Lambda$CDM are statistically equivalent based on AICc ($\Delta$AICc$<4$), when the sample are 
treated separately, but show a strong evidence against ($6<{\rm BIC}<10$) the scenario when the 
joint analysis is applied. On the other hand, although AICc suggests that 
$\Lambda \neq 0$ model and $\Lambda$CDM are statistically equivalent in the 
joint analysis, BIC indicates that there is a strong evidence against   
the candidate model. 
Additionally, for the two models we find 
that the $\Lambda=0$ case is preferred by separate datasets, while the 
$\Lambda\neq 0$ case is statistically preferred for the combined data analysis.

For completeness, we additionally calculate
the Deviance information 
criterion \citep[DIC,][]{Spiegelhalter:2002, Kunz:2006, Liddle:2007}. This is 
defined as  ${\rm DIC} = D(\bar{\theta}) + 2p_D$,
where $D(\bar{\theta}) = \chi^2(\bar{\theta})$ is the Bayesian deviation,
$p_D = \bar{D}(\theta) - D(\bar{\theta})$ is the Bayesian complexity, which 
represents the number of effective degrees of freedom, and $\bar{\theta}$ is the 
mean value of the parameters. 
The advantage of DIC is its use of the full 
log-likelihood sample instead of only the maximum log-likelihood (or minimum 
$\chi^2$) as AICc and BIC do. Based on the Jeffreys scale \citep{Jeffreys:1961}, 
for $\Delta{\rm DIC}<2$ both models are statistical equivalent. In contrast, 
$2<\Delta {\rm DIC}<6$ suggests a moderate tension between models, being the one 
with lower value of DIC the best one, and $\Delta {\rm DIC}> 10$ implies a 
strong tension between the two models. We find that the $\Lambda \neq 0$ case 
and $\Lambda$CDM scenario are statistical equivalent for BAO, they have a 
moderate tension for OHD and SLS, and a strong tension for HIIG and SNIa. On the 
other hand, the $\Lambda \neq 0$ case and $\Lambda$CDM are statistical 
equivalent for OHD, BAO, HIIG, and SNIa. In summary, we confirm the results 
obtained for the Joint analysis by AICc and BIC for both $\Lambda\neq 0$ and 
$\Lambda=0$ models. It is worth to mention that  when a posterior distribution 
presents a bimodal shape or is asymmetric for a parameter, $p_D$ yields negative 
values and thus DIC may not be a good criterion. This situation is mainly 
presented for $\beta$ in the $\Lambda \neq 0$ case in separate datasets.

As a next step, we use the constraints from the joint analysis to reconstruct 
the  three cosmographic parameters, namely the Hubble, $H(z)$, 
the  
deceleration, $q(z)$,  and jerk, $j(z)$, parameters according 
to (\ref{q}), (\ref{j}). The cosmic 
evolution of   parameters   is shown in Fig.
\ref{fig:Hz_and_qz}.  Thus, we report the current values of 
$q_0=-0.610^{+0.028}_{-0.035}$ ($-0.708^{+0.016}_{-0.016}$) for  the 
deceleration parameter, and  $j_0 = 1.041^{+0.051}_{-0.036}$ 
($1.137^{+0.014}_{-0.013}$) for the jerk parameter 
for the $\Lambda \neq 0$ 
($\Lambda=0$) scenario.
Furthermore, the transition 
redshift 
between the deceleration 
and the acceleration stages is estimated to be 
$z_T=0.715^{+0.042}_{-0.041}$ ($0.652^{+0.032}_{-0.031}$), which is in agreement 
with the one obtained by $\Lambda$CDM as shown in Fig. \ref{fig:Hz_and_qz}. 
Note that the jerk 
parameter evolution reveals the  dynamical  equation of 
state of the effective dark energy.

\begin{figure}
    \centering
    \includegraphics[width=0.45\textwidth]{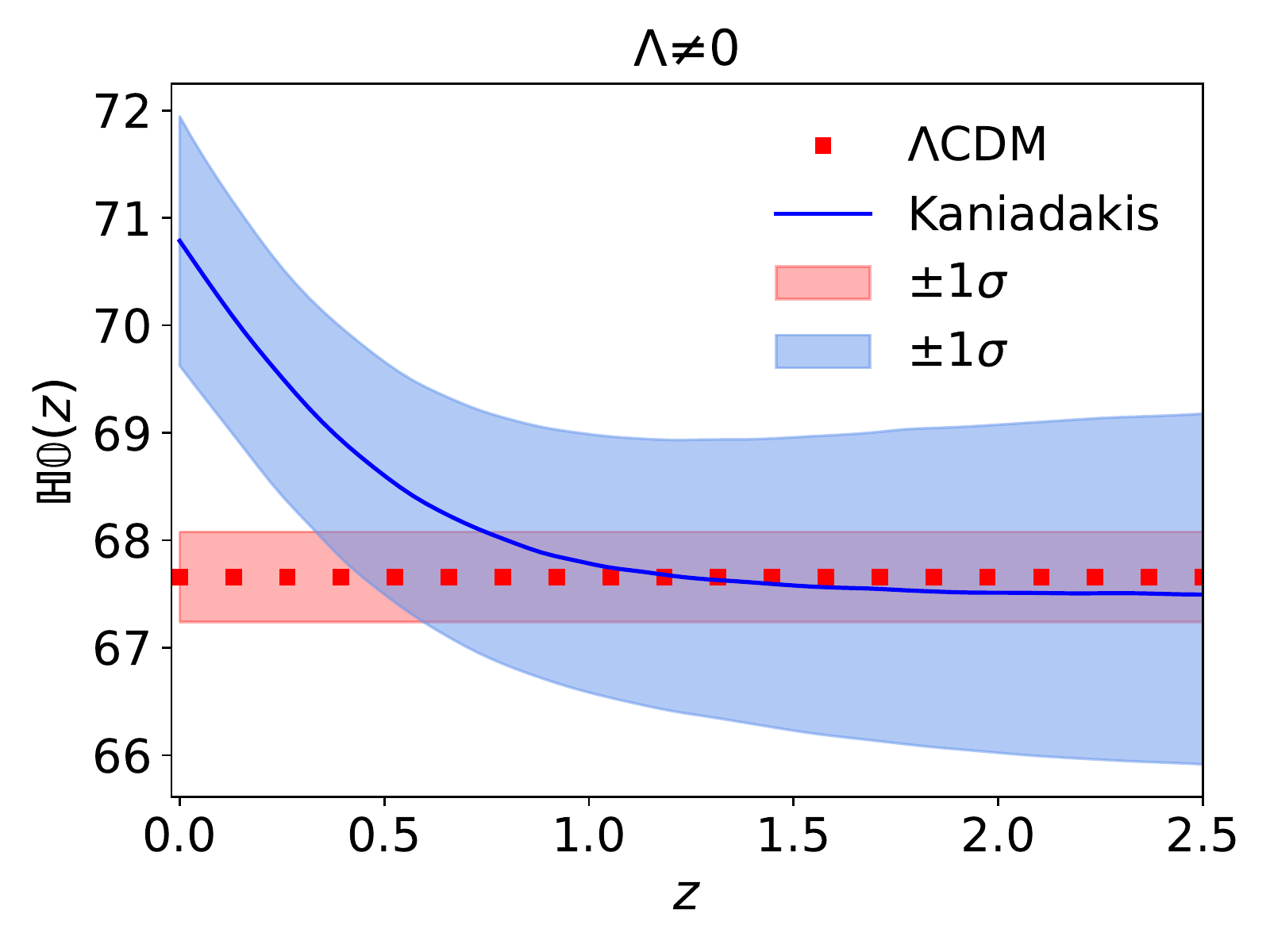}
    \includegraphics[width=0.45\textwidth]{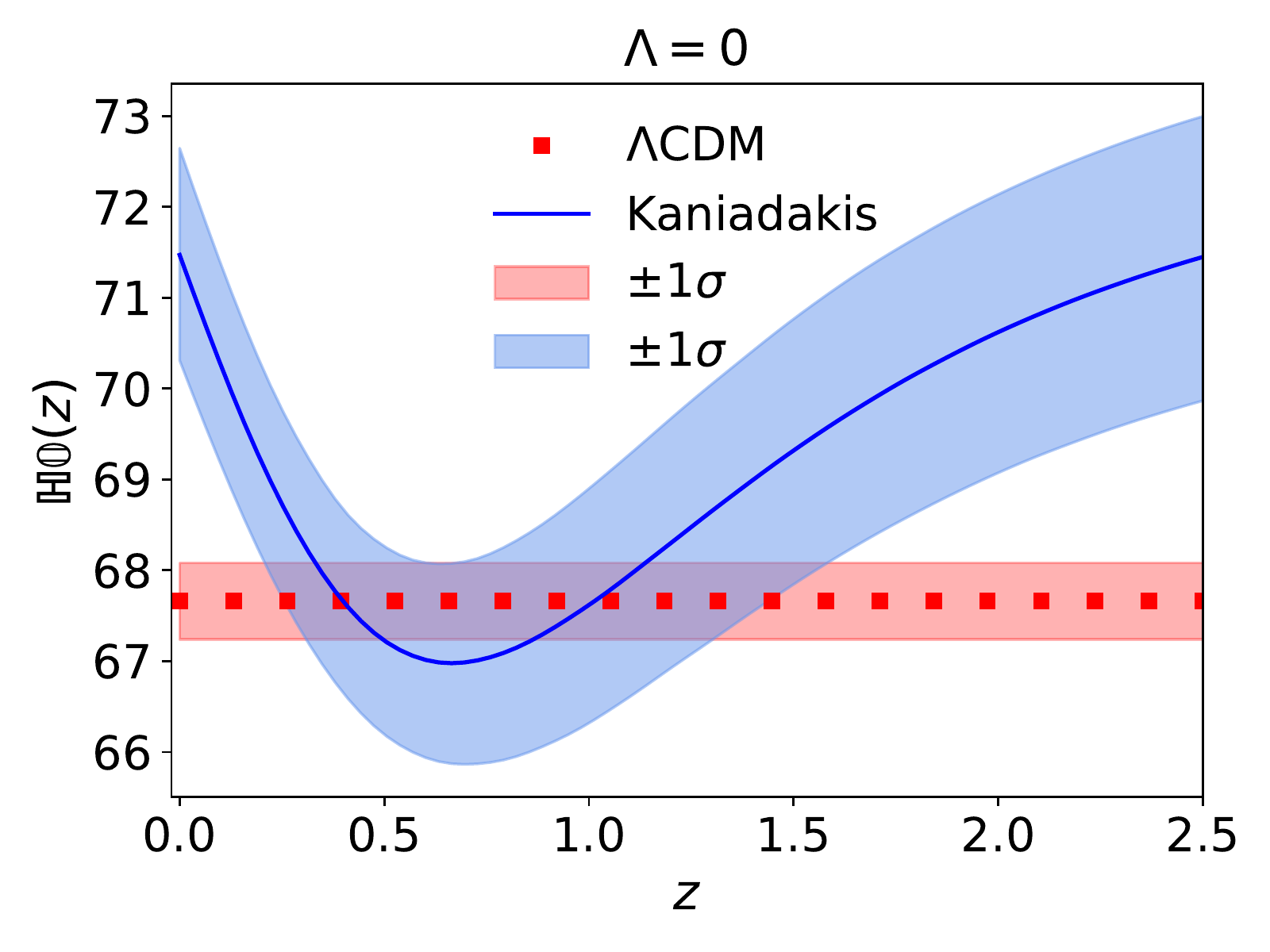}
    \caption{The $\mathbf{\mathbb{H}}0(z)$ diagnostic for Kaniadakis horizon 
entropy cosmology with 
$\Lambda \neq 0$ (upper panel) and $\Lambda = 0$ (lower panel).  We have used 
the bound obtained from the joint 
analysis, and the shaded regions denote the uncertainties at $1\sigma$. 
For completeness, the red square-points 
represent the results of  $\Lambda$CDM cosmology  with $h=0.6766$ and
$\Omega_m^{(0)}=0.3111$ 
\citep{Planck:2018vyg}.}
    \label{fig:H0diagnostic}
\end{figure}

Finally, 
to investigate in more detail  the Hubble tension, we apply a 
new diagnostic, called $\mathbf{\mathbb{H}}0(z)$ diagnostic, defined by 
\citep{H0diagnostic:2021}
\begin{equation}
    \mathbf{\mathbb{H}}0(z) = \frac{H(z)}{E_{\rm \Lambda CDM}(z)}\,,
\end{equation}
where $H(z)$ is the Hubble function evolution in a given cosmological scenario  
alternative   to $\Lambda$CDM, 
and $E_{\rm \Lambda CDM}(z)$ is the 
dimensionless Hubble parameter of $\Lambda$CDM paradigm. This diagnostic 
measures a possible deviation of $H_0$ from its  $\Lambda$CDM value. Concerning 
 a flat $\Lambda$CDM, 
 a non-constant path of $\mathbf{\mathbb{H}}0(z)$ 
within error bars suggests a modification of the Planck-$\Lambda$CDM 
scenario. In Fig. \ref{fig:H0diagnostic} we depict the obtained  
results. 

As we observe, there is an agreement within $1\sigma$ between flat-$\Lambda$CDM cosmology and  Kaniadakis cosmology for $z\gtrsim 0.7$ in the $\Lambda\neq 0$ case, and for $0.7 \lesssim z \lesssim 1.3$ in the $\Lambda= 0$ case.
Additionally, it is interesting that the current value 
$\mathbf{\mathbb{H}}0(z=0)$ for both models are consistent with the one obtained 
by SH0ES \citep{Riess:2019}, and that the $\Lambda\neq0$ model has a trend to the Planck 
value in the past \citep{Planck:2018vyg}. This is another verification that the 
scenario of Kaniadakis horizon entropy cosmology may offer an alleviation to 
the $H_0$ tension. 
Nevertheless,  to further investigate whether both Kaniadakis models can 
alleviate the Hubble tension, a parameter estimation using the linear 
perturbation equations together with CMB data should be performed.

We close this section by investigating one important process in every 
cosmological scenario: the Big Bang Nucleosynthesis (BBN), since
  the production of light elements in 
the early Universe  can be 
affected in non-standard cosmologies \citep{Pospelov:2010ARNPS, 
Barrow:2021PhLB}. Considering that the freeze-out of the light elements occurs 
when the weak interaction rates are lower than $H(z)$, a simple test to 
guarantee that the BBN is not spoiled is to require that the deviation $\delta 
H(z)$ with respect to the standard Hubble expansion rate at the BBN epoch should 
be small. Although in the Friedmann equations mentioned above we have not 
included a radiation component, this can be added and  we can perform the 
  analysis by expanding $E_4$ around $\beta=0$ in Eq. 
\eqref{E4} (resp. Eq. \eqref{eqN340}) and neglecting the fourth order error 
terms, resulting to
\begin{small}
\begin{align}
& E(z)= \left\{
\begin{array}{c}
\underbrace{ \sqrt{\Omega_m^{(0)}  (z+1)^{{3 }}+ 
\Omega_\Lambda^{(0)}}}_{\text{value from}\; \Lambda {\rm CDM}}+ \underbrace{\frac{\beta ^2}{4}
  \left[\Omega_m^{(0)}  (z+1)^{{3 }}+ 
\Omega_\Lambda^{(0)}\right]^{-3/2}}_{\text{correction}}, \; \Lambda \neq 0 \\
 \underbrace{ \sqrt{\Omega_m^{(0)} } (z+1)^{{3/2}}}_{\text{value from}\; \Lambda {\rm CDM}}+ \underbrace{\frac{\beta ^2}{4}
  \left[\Omega_m^{(0)}\right]^{-3/2}  (z+1)^{{-1/2}}}_{\text{correction}}, \; \Lambda=0
\end{array}\right..
\label{EQ_53}
\end{align}
\end{small}
Hence, we can study both Kaniadakis models ($\Lambda \neq 0$ and $\Lambda=0$) 
at $z\sim10^{10}$ (approximately BBN era). 
We find that  for $\Lambda\neq 0$,  the model is consistent with the Big Bang 
Nucleosynthesis (BBN) constraints, since  the correction term at 
$z\sim10^{10}$ is of the order of $\sim 10^{-49}$, dominating the standard 
cosmology and not producing significant effects in the formation of light 
elements.
In the case $\Lambda=0$, the correction is larger, and calculations at 
$z\sim10^{10}$ are of the order $\sim10^{-5}$. However, such corrections are 
still subdominant, allowing the production of light elements. A further 
analysis could be performed following 
\citet{Capozziello:2017bxm,Barrow:2021PhLB,Asimakis:2021yct}.

\section{Dynamical system and stability analysis} \label{sec:SA}

In this section we perform a full dynamical system analysis in  order to 
investigate the  global dynamics of cosmological scenarios, and obtain 
information on the Universe evolution independently of the initial 
conditions. 
In the dynamical system formulation, one   starts from local analysis  of the differential equation $\mathbf{x}'(\tau)={\bf 
X}(\mathbf{x})$, where $\mathbf{x}$ is the state vector, and $\tau$ a convenient time variable, near an equilibrium point $\mathbf{x}=\bar{\mathbf{x}}$, and  
progressively extends    the   investigated regions of the phase  and of the 
parameter space.  Assuming that the vector field ${\bf 
X}(\mathbf{x})$ has continuous partial
derivatives, the process of determining the local behavior is based on the 
linear approximation of the vector field ${{\bf X}(\mathbf{x}) \approx {\bf 
DX}(\bar{\mathbf{x}})(\mathbf{x}-\bar{\mathbf{x}})}$ where ${\bf DX}(\bar{\mathbf{x}})$ is the Jacobian of the  vector field at the 
equilibrium point $\bar{\mathbf{x}}$, which is referred to as the {\it 
linearization  of the dynamical equations  at the equilibrium point}. In this neighborhood we acquire the system $\mathbf{x}'(\tau)={ {\bf 
DX}(\bar{\mathbf{x}})(\mathbf{x}-\bar{\mathbf{x}})}$.  Each of 
the equilibrium points can be classified according to the real parts of the eigenvalues of  ${\bf 
DX}(\bar{\mathbf{x}})$ (if none of these are zero).  Thus, this approach provides a general description of 
the phase space of all   possible solutions of the system, their equilibrium 
points and stability, as well as the asymptotic solutions  
\citep{Ellis,Ferreira:1997au,Copeland:1997et,Perko,Coley:2003mj,Copeland:2006wr,
Chen:2008ft,Cotsakis:2013zha,Giambo:2009byn,Papagiannopoulos:2022ohv}. If some 
real parts of the eigenvalues are zero,  the equilibrium point is nonhyperbolic, 
and the analysis through linearization fails. Then, we use numerical tools for 
the analysis. 

In the following subsections we perform the global dynamical system analysis for the two cases, namely  $\Lambda \neq 0$ and  $\Lambda = 0$.

  \subsection{Case I: $\Lambda \neq 0$}

Defining the dimensionless variables $\theta, T$   as
\begin{align}
\theta=   \arctan \left( 1- \frac{8 \pi G \rho_{DE}}{3 H^2 }\right), \; \theta\in \left[-\frac{\pi}{2},\frac{\pi}{2}\right],\quad     T= \frac{H_0}{H+ H_0}, 
\label{Tthetavar}
\end{align}
with 
\begin{equation}
\Omega_m:= \frac{8 \pi G \rho_m}{3 H^2 }  = \tan \left(  \theta \right),
\end{equation}
then equation \eqref{gfe2} becomes
\begin{small}
\begin{align}
\frac{\Lambda}{3 H_0^2 }= \frac{(1-T)^2  \left\{\cosh \left[\frac{T^2 \beta 
}{(1-T)^2}\right]-\tan (\theta )\right\}}{T^2}-\beta  \text{shi}\left[\frac{T^2 
\beta
   }{(1-T)^2}\right].
\end{align}
\end{small}
It proves convenient to 
introduce the new time derivative as 
\begin{align}
    f^{\prime} \equiv \frac{d f}{d \tau}= \frac{\cosh \left(\frac{\pi  K}{G 
H^2}\right)}{H}  \dot{f} .
\end{align}
Therefore, we finally extract the dynamical system 
\begin{align}
& {\theta}^{\prime}(\tau)= 3 \sin (\theta ) \left\{\sin (\theta )-\cos (\theta 
) \cosh \left[\frac{T^2 \beta
   }{(1-T)^2}\right]\right\}, \label{eq4.10}\\
& T^{\prime}(\tau)=  \frac{3}{2} (1-T) T \tan (\theta). \label{eq4.9}
   \end{align}
Note that this system diverges  at $T=1$ and at $\theta=\pm\pi/2$.
 
 Lastly, the deceleration parameter (\ref{q}) is written as
\begin{equation}
    q:= -1-\frac{\dot{H}}{H^2}=  -1 + \frac{3}{2} \tan (\theta ) 
\text{sech}\left[\frac{\beta  T^2}{(1-T)^2}\right]. 
\end{equation}

Note that, for an expanding universe ($H>0$), we have that $T\in[0,1]$, while  
$\theta$ 
is a periodic coordinate with period $\pi$,  and thus we can set 
$\theta\in\left[-{\pi}/{2}, {\pi}/{2}\right]$ (modulo a periodic shift $c \pi, 
\; c\in\mathbb{Z}$). Moreover, the physical condition $0\leq \Omega_m\leq1$ 
implies that the region of physical interest is $\theta \in[0, \pi/4]$ (modulo 
a periodic shift $c \pi, \; c\in\mathbb{Z}$). The non-physical region 
$\Omega_m>1$ is $\theta\in(\pi/4,  \pi/2]$ (modulo a periodic shift $c \pi, 
\; c\in\mathbb{Z}$). 
Hence, we have obtained a global phase-space formulation.
 For the representation of the flow of \eqref{eq4.10} and \eqref{eq4.9}, we integrate in the variables $T, \theta$ and project in a compact set
using the ``cylinder-adapted'' coordinates 
\begin{equation}
\label{cylinder}
 \mathbf{S}: \begin{cases}
x = \cos ( \theta),\\ 
y = \sin ( \theta),\\ 
z = T, 
\end{cases}
\end{equation}
with $0\leq T\leq 1, \theta \in [-\pi, \pi]$,
with inverse 
$ 
\theta =  \arctan \left({y}/{x}\right)$, and $ T =z$. 
Thus, the region of physical interest is  $\theta \in [0, \pi/4]$, 
modulo 
a periodic shift $c \pi, \; c\in\mathbb{Z}$.

\begin{table}  
    \centering
    \begin{tabular}{|c|c|c|c|c|}\hline
   Label & $\theta$ &  $T$& Existence & Stability  \\\hline
   $dS_{+}$ &   $ 2 \pi  c_1$  & arbitrary & $c_1\in \mathbb{Z}$      & stable  \\ 
   $dS_{+}^{(0)}$     &    $ 2 \pi  c_1 $ & $0$ & $c_1\in \mathbb{Z}$  & stable  \\
   $dS_{+}^{(1)}$    &    $ 2 \pi  c_1 $   & $1$ & $c_1\in \mathbb{Z}$ & stable  \\
 $dS_{-}$    &    $\pi(1 + 2 c_1) $  & arbitrary & $c_1\in \mathbb{Z}$  & stable  \\
   $dS_{-}^{(0)}$    &    $\pi(1 + 2 c_1) $ & $0$  & $c_1\in \mathbb{Z}$  & stable  \\
   $dS_{-}^{(1)}$   &    $\pi(1 + 2 c_1) $  & $1$ & $c_1\in \mathbb{Z}$  & stable  \\
 $M_{-}^{(0)}$ &  $2 \pi  c_1-\frac{3 \pi }{4} $  & $0$ & $c_1\in \mathbb{Z}$ & unstable \\
  $M_{+}^{(0)}$&  $2 \pi  c_1+\frac{\pi }{4} $  & $0$  & $c_1\in \mathbb{Z}$  & unstable \\
  $M_{-}^{(1)}$ &  $2 \pi  c_1-\frac{3\pi }{4} $   & $1$ & $c_1\in \mathbb{Z}, \beta=0$ & saddle \\
  $M_{+}^{(1)}$ &  $2 \pi  c_1+\frac{\pi }{4} $ &  $1$ & $c_1\in \mathbb{Z}, \beta=0$ & saddle  \\\hline
    \end{tabular}
    \caption{The equilibrium points of the dynamical system  \eqref{eq4.10} 
and \eqref{eq4.9} of Kaniadakis horizon entropy cosmology with $\Lambda\neq0$. 
We use  $dS$ to denote the de Sitter, dark-energy dominated solutions, and $M$ to denote the matter-dominated ones.
}
    \label{Eigen-1}
\end{table}
We proceed by extracting the equilibrium points and 
characterizing their stability. There  are two equivalent hyperbolic 
equilibrium points $M_{\pm}$ for which $q=1/2$, 
i.e. they are 
associated with matter domination, and two equilibrium points 
$dS_{\pm}$ corresponding to dark-energy dominated de-Sitter solutions for which 
$q = -1$.  The equilibrium points of the dynamical system  \eqref{eq4.10}
and  \eqref{eq4.9} of Kaniadakis horizon entropy cosmology with $\Lambda\neq0$ are presented in Table \ref{Eigen-1}. 
  \begin{figure}
    \centering
    \includegraphics[scale=0.6]{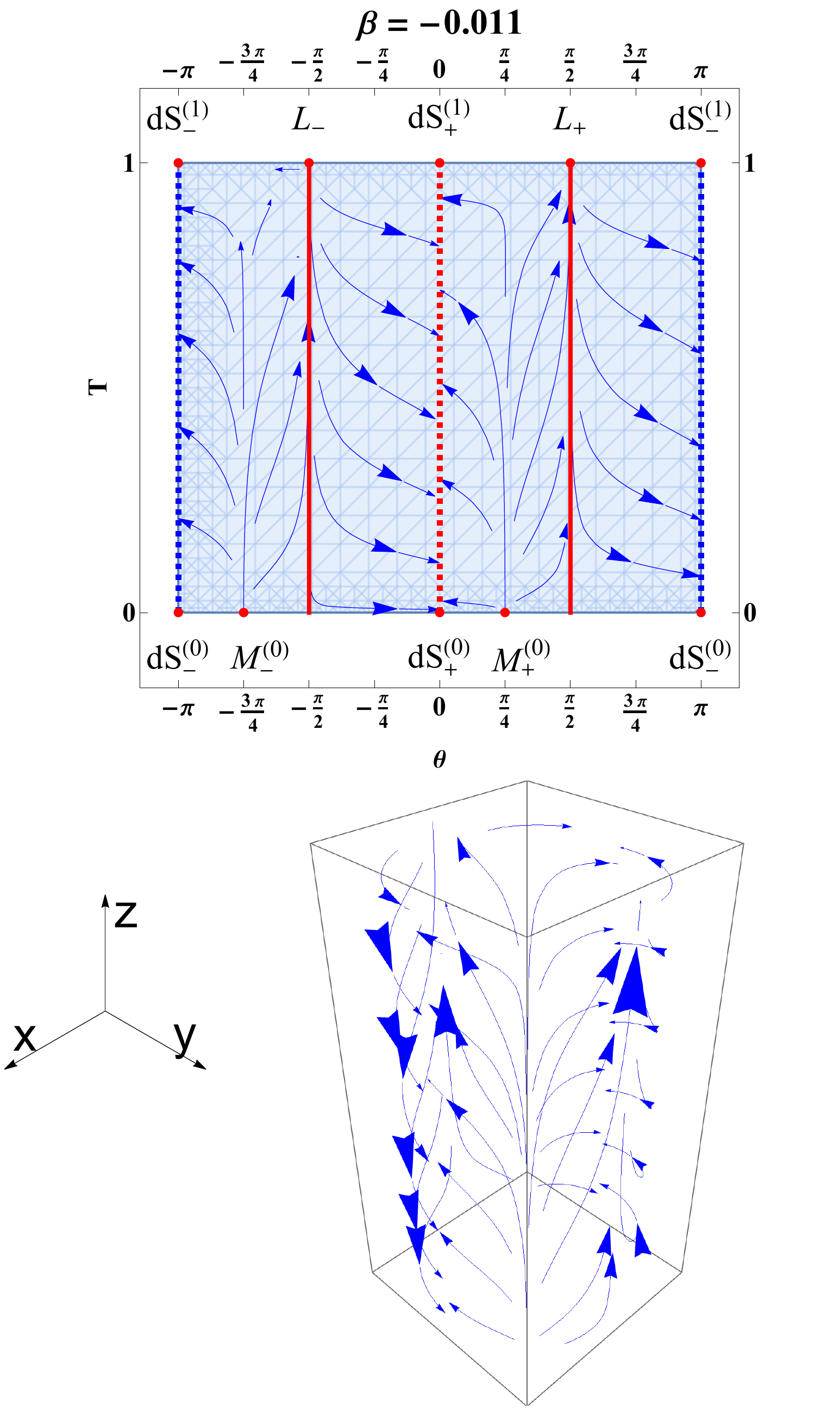}
    \caption{Phase-space plot of the dynamical system \eqref{eq4.10} 
and   \eqref{eq4.9} of Kaniadakis horizon entropy cosmology with $\Lambda\neq0$, 
for the best-fit value  of Kaniadakis  parameter obtained by the observational 
analysis, namely for $\beta=-0.011$. Upper panel:
 unwrapped solution space. Lower panel: projection over the 
cylinder $\mathbf{S}$ defined in Cartesian coordinates $(x,y,z)$  through 
\eqref{cylinder}. At late times the Universe results 
in a dark-energy dominated, de Sitter solution, while 
 the   past attractor is the matter-dominated epoch.}
    \label{fig:my_label2}
\end{figure}
In Fig. \ref{fig:my_label2} we display an unwrapped solution space of the system \eqref{eq4.10} and \eqref{eq4.9} (upper 
panel), and the projection over the cylinder $\mathbf{S}$, defined in Cartesian  coordinates $(x,y,z)$  by \eqref{cylinder}, for the best fit value $\beta=-0.011$ obtained through the observational analysis.  
For the points that are non-hyperbolic, their stability is analyzed 
numerically.   The two dashed lines, indicated by $dS_{-}$ (blue) and  
 $dS_{+}$ (red),  are the late-time de Sitter attractors.  The early-time attractors 
are  $M_{\pm}^{(0)}$ for which $q=1/2$, 
and they  correspond to 
matter-dominated solutions.  Hence, 
 at late times the Universe results 
in a dark-energy dominated solution, while 
 the   past attractor of the Universe is the matter-dominated epoch. At 
the intersection of the invariant set $T=1$ with the singular lines $\theta=\pm 
\pi/2$ we obtain the equilibrium points $L_{\pm}$. Considering that equations 
\eqref{eq4.10} and \eqref{eq4.9} diverge at $L_{\pm}$, we should 
introduce suitable variables for the analysis.

For the analysis at $T=1$  it proves convenient to define the variable 
\begin{equation}
    \Phi=  \left\{1+\exp \left[\frac{|\beta|
   T^2}{(1-T)^2}\right]\right\}^{-1}, \Phi \in[0,1], 
\end{equation}
as well as   the time rescaling 
\begin{equation}
        f^{\prime} \equiv  \frac{d f}{d \zeta}=    (1-\Phi)^2 \frac{d f}{d 
\eta}=     \frac{\tanh ^2\left(\frac{\pi  
|K|}{2 G H^2}\right)+1}{4 H} \dot{f}.
\end{equation}
Hence, using also the  variable  $\theta$ from (\ref{Tthetavar}),  
we finally  
obtain the autonomous system
\begin{align}   &  \theta^{\prime}(\zeta)= -\frac{3}{2} \sin (\theta ) \left[2 (\Phi -1) \Phi  
(\sin (\theta )+\cos (\theta ))+\cos (\theta )\right], \label{4.39}\\
   & \Phi^{\prime}(\zeta)=   3 (1-\Phi)^2 \Phi ^2 \tan (\theta ) \ln 
\left[\Phi/(1-\Phi)\right]. \label{4.38}
\end{align}

\begin{figure} 
    \centering
   \includegraphics[scale=0.9]{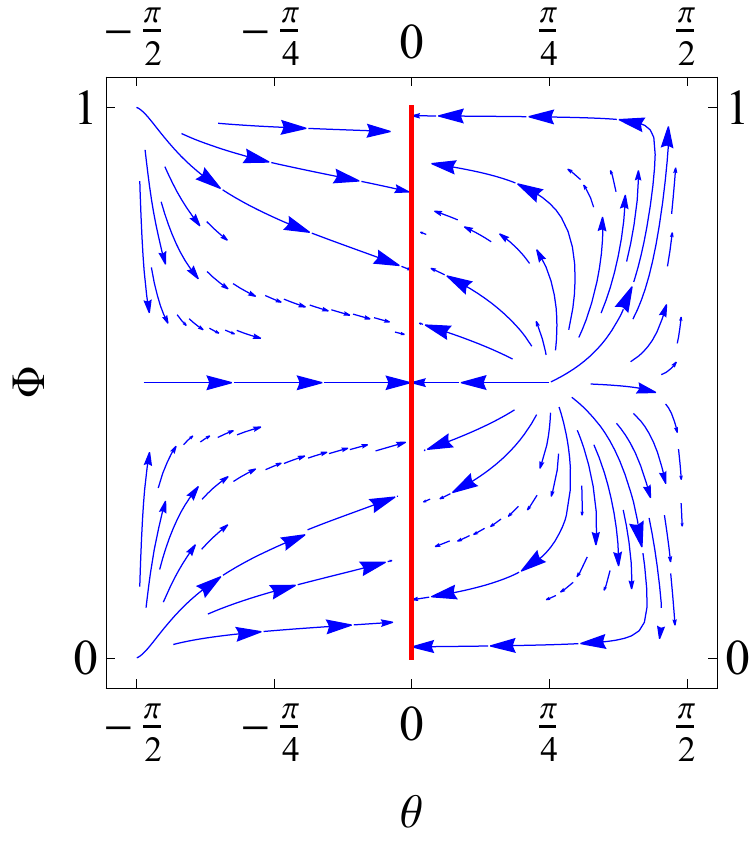}
    \caption{Phase-space plot of the system \eqref{4.39}-\eqref{4.38} of 
Kaniadakis horizon entropy cosmology with $\Lambda\neq0$,  for dust matter. The 
late attractor corresponds to $ \theta=0$, and thus to a dark-energy 
domimated solution with 
$\Omega_{DE}= 1$.   }
    \label{T1}
\end{figure}
In Fig. \ref{T1} we depict the phase-space flow     of the system 
\eqref{4.39}-\eqref{4.38}. Asymptotically, $\theta\rightarrow 0$ and $\Phi$ tends to a constant $\Phi_0$.  
Therefore, the late attractor corresponds to the dark-energy dominated solution 
with   $\Omega_{DE}= 1$. The current values $ \Phi_{0}= 1/(e^{\left| \beta \right| }+1)$, $\theta_0=\text{arctan}\left(\Omega_m^{(0)}\right)$ leads to the de Sitter solution  $a(t)=e^{H_0 (t-t_U)}$.

\subsubsection{Heteroclinic sequences}
\label{section4.1.1}

In the phase portrait of a  dynamical system, a heteroclinic orbit is a 
path in phase space that joins two different equilibrium points. If the 
equilibrium points at the start and end of the orbit are the same, the orbit is 
a homoclinic orbit \citep{Guckenheimer}.

\begin{figure} 
    \centering
   \includegraphics[scale=0.3]{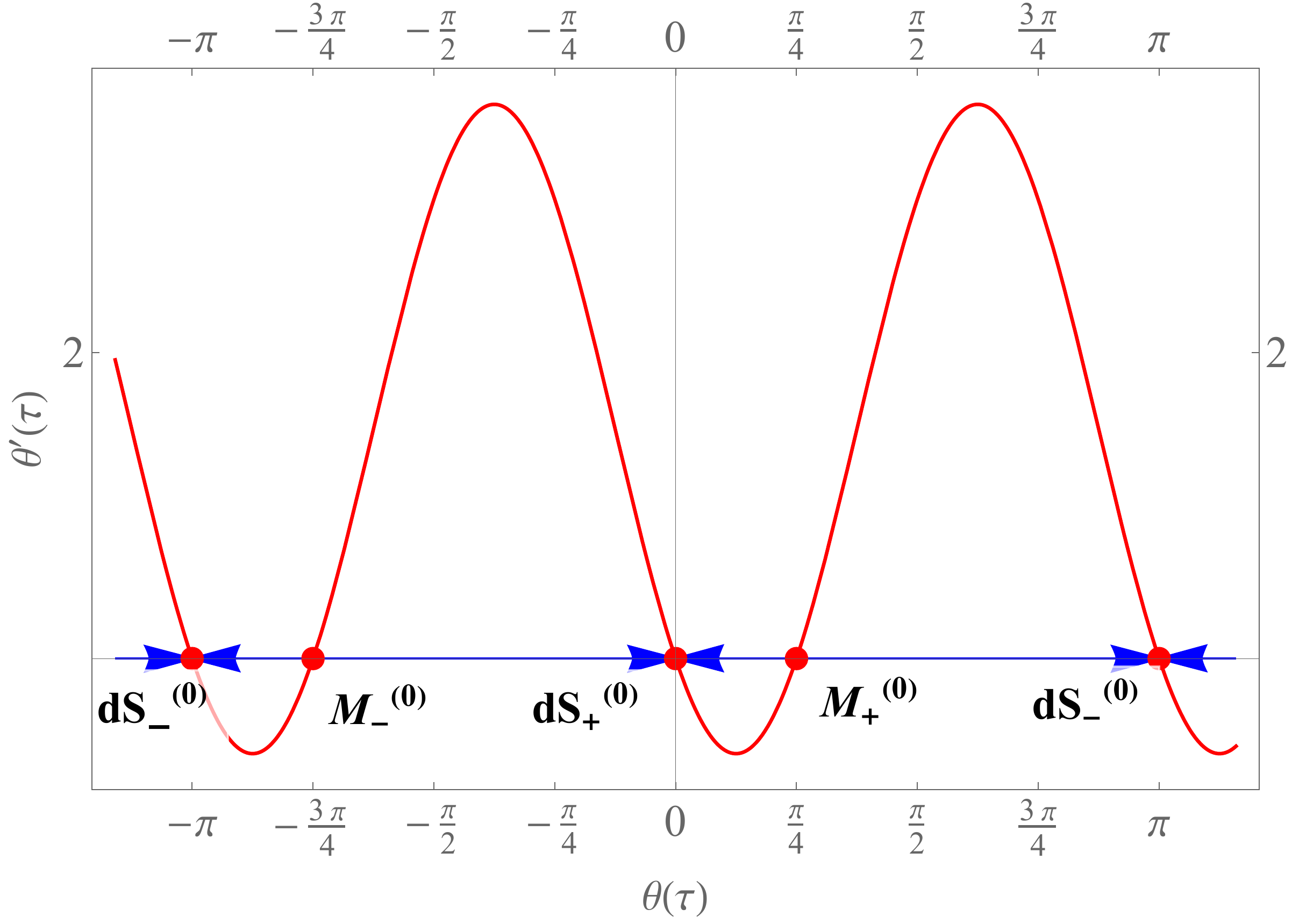}
    \caption{Phase-space diagram of the one-dimensional dynamical system \eqref{one-FIG4A} of 
Kaniadakis horizon entropy cosmology with $\Lambda\neq 0$,  for dust matter and  any value $\beta$. }
    \label{FIG4A}
\end{figure}

From the above analysis we can see that the invariant sets $T=0$ and 
$T=1$ are of interest in the determination of possible heteroclinic sequences. 
The direction of the flow can be determined by considering the monotonic 
function 
 \begin{equation}
   M_1= \frac{T}{1-T},\quad M_1'(\tau)=\frac{3 \tan (\theta )}{2} M_1.
 \end{equation}
In if $\tan(\theta)<0$, the orbits move from $T=1$ to $T=0$, and if  
$\tan(\theta)>0$, the orbits move from $T=0$ to $T=1$. In the invariant manifold 
$T=0$ ($H\rightarrow \infty$) the dynamics is given by the one-dimensional 
flow 
 \begin{equation}
 \label{one-FIG4A}
  {\theta}^{\prime}(\tau)= 3 \sin (\theta ) \left(\sin (\theta )-\cos (\theta 
) \right).
 \end{equation}
In Fig.  \ref{FIG4A} we present the one-dimensional dynamical system 
\eqref{one-FIG4A}, in which we can see the heteroclinic sequences $M_+^{(0)} 
\to d S_{+}^{(0)}$
and $M_-^{(0)} \to d S_{-}^{(0)}$.
Similarly, analyzing the one-dimensional flow in the invariant 
set $T=1$, that corresponds to $\Phi=0$, we find that the dynamics on this 
invariant set is given by the one-dimensional dynamical system 
\begin{equation}
\label{one-FIG4B}
 \theta^{\prime}(\zeta)= -\frac{3}{2} \sin (\theta ) \cos (\theta),
 \end{equation}
  \begin{figure} 
    \centering
   \includegraphics[scale=0.3]{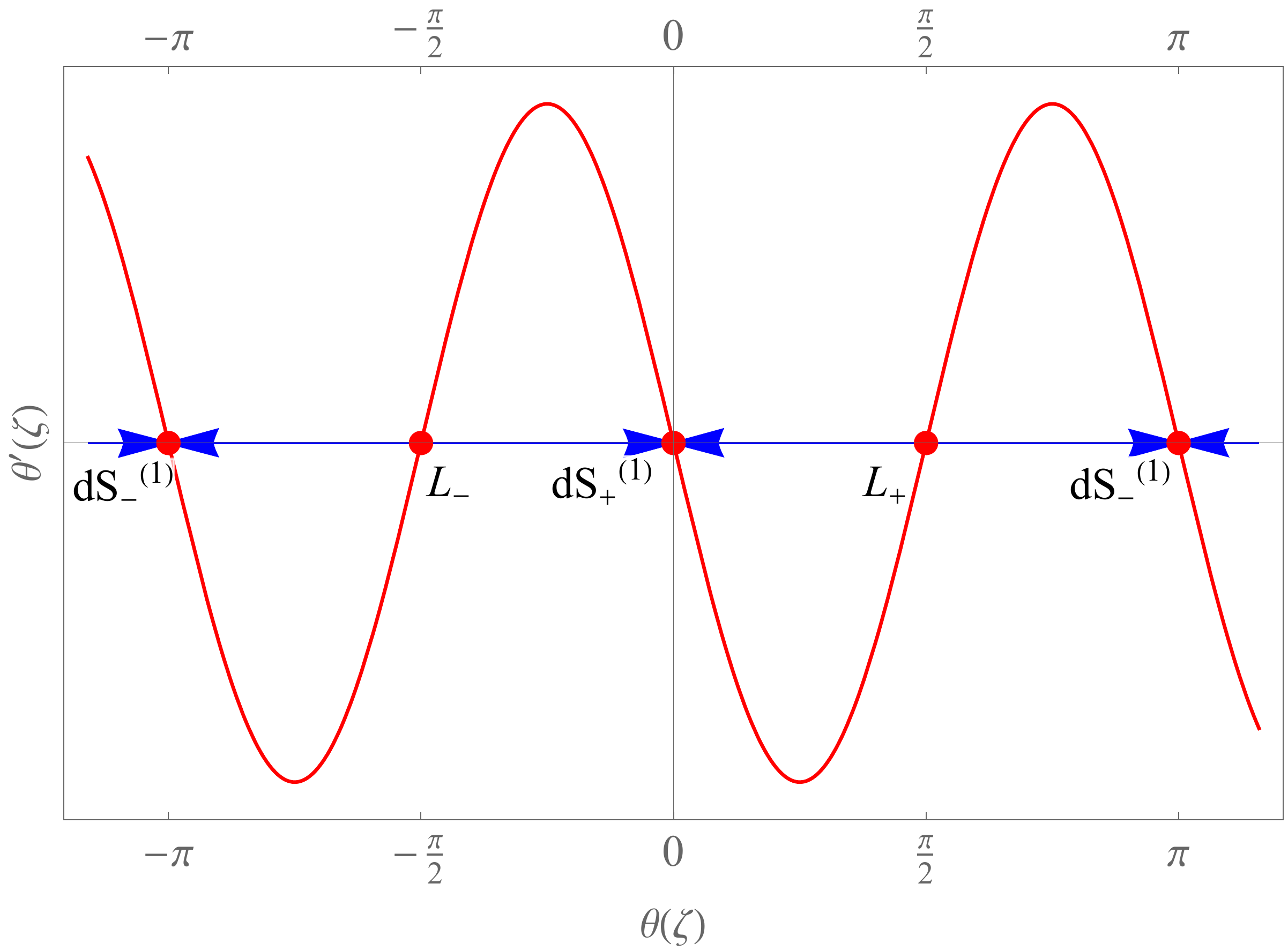}
    \caption{Phase-space diagram of the one-dimensional dynamical system \eqref{one-FIG4B} of 
Kaniadakis horizon entropy cosmology with $\Lambda\neq 0$,  for dust matter and  any value $\beta$. }
    \label{FIG4B}
\end{figure}
 which has a behavior shown in Fig. \ref{FIG4B}, where  the 
heteroclinic sequences $L_+ \to d S_{+}^{(1)}$ and $L_- \to d S_{-}^{(1)}$ are presented.
Finally, to find heteroclinic sequences $M_\pm^{(0)}  \to d 
S_{\pm}^{(1)}$, the intersection of the unstable manifold of $M_\pm^{(0)}$ with 
the stable manifold of $d S_{\pm}^{(1)}$ should be analyzed. Since the former 
is $\mathbb{R}^2$, then it is required to examine the stable manifold of  $d 
S_{\pm}^{(1)}$. 
This is given locally by the graph 
\begin{equation}
    \left\{(\Phi, \theta)\in  \mathbb{R}^2: \Phi= h(\theta), h(0)=0, h'(0)=0 
\right\}, |\theta|<\delta,
\end{equation}
for $\delta>0$ suitably small. 
By the invariance of the stable manifold  we obtain the quasilinear 
differential equation for $h$ given by 
\begin{align}
& \frac{3}{2} \sin (\theta ) h'(\theta ) (2 (h(\theta )-1) h(\theta ) (\sin 
(\theta )+\cos (\theta ))+\cos (\theta )) \nonumber \\
& +3
   (h(\theta )-1)^2 h(\theta )^2 \tan (\theta ) \ln \left(\frac{h(\theta 
)}{1-h(\theta )}\right)=0. 
\end{align}
Introducing the ansatz 
$
    h(\theta)= a_1 \theta^2 + a_2 \theta^3 + a_3 \theta^4 + \ldots,
$
we obtain $a_i=0$ at any order. Therefore, the dynamics at the stable manifold 
of  $d S_{+}^{(1)}$ is given by   equation \eqref{one-FIG4B}. Then, it is easy 
to construct  heteroclinic sequences $M_+^{(0)} \to d S_{+}^{(1)}$ which pass 
near the singularity $L_+$ by assuming for instance the initial value $(\Phi, 
\theta)=(\varepsilon, \pi/4)$, $\varepsilon \approx 0$ and evolving the system 
back and forward in $\zeta$. Similar arguments can be used to construct 
heteroclinic sequences $M_-^{(0)} \to d S_{-}^{(1)}$, which pass near the 
singularity $L_-$, with the initial value $(\Phi, \theta)=(\varepsilon, 
-3\pi/4)$, $\varepsilon \approx 0$.

Summarizing, for $0\leq \theta \leq \pi/2$ (the physical region is $0\leq \theta 
\leq \pi/4$), there exists the heteroclinic sequences $M_+^{(0)}\; 
(\Omega_m\rightarrow 1, H\rightarrow \infty) \to L_+ (\Omega_m\rightarrow 
+\infty, H\rightarrow 0) \to d S_{+}^{(1)}  (\text{de Sitter}, 
\Omega_m\rightarrow 0, H\rightarrow 0)$ and $M_+^{(0)} \to d S_{+}^{(0)} 
(\text{de Sitter}, \Omega_m\rightarrow 0, H\rightarrow \infty)$, and in the 
region $-\pi \leq \theta \leq -\pi/2$ (the physical region is $-\pi\leq \theta 
\leq - 3\pi/4$), there exists the heteroclinic sequences $M_-^{(0)}  
(\Omega_m\rightarrow 1, H\rightarrow \infty) \to L_- (\Omega_m\rightarrow  
-\infty, H\rightarrow 0) \to d S_{-}^{(1)}  (\text{de Sitter}, 
\Omega_m\rightarrow 0, H\rightarrow 0)$ and $M_-^{(0)} \to d S_{-}^{(0)}  
(\text{de Sitter}, \Omega_m\rightarrow 0, H\rightarrow \infty)$.

\subsubsection{Bounce and a turnaround}

Another interesting cosmological possibility is   the possible 
existence of a bounce and a turnaround 
\citep{Saridakis:2007cf,Cai:2012ag,Zhu:2021whu}. Let us assume that, for the 
state vector $(a, H, R)$, the field equations can be written as  
\begin{align}
   & \dot{a}= a H, \label{eq_a}\\
   & \dot{H}= \frac{1}{6}\left(R - 12 H^2\right),\label{eq71} \\
   & \dot{R}= g(a, H, R),\label{eq_g}
\end{align}
such that the function $g(a, H, R)$ satisfies $g(a, H, R)= - g(a, -H, R)$. 
Hence, the system \eqref{eq_a}, \eqref{eq71} and \eqref{eq_g} is invariant under 
time inversion $t \mapsto -t$ if also $H \mapsto -H$ and $R \mapsto R$, and by 
definition $a\geq 0$. Those solutions can be related to symmetric cyclic 
solutions with respect to the origin,  chosen to correspond to the 
possible bounce point  $t_{\text{bounce}}=0$. Therefore, if the bounce 
exists, the system  \eqref{eq_a}, \eqref{eq71} and \eqref{eq_g}  is a reversible 
system in the sense that it has a reversing symmetry under time inversion.

Let us 
consider the simplest case where there is exactly one bouncing and exactly one 
turnaround point. Note that both at the bounce and turnaround points we have 
$H=0$. In this case, the line connecting these points and corresponding to $H=0$ 
defines a plane that separates all   points on the trajectory in this phase 
space to the ones corresponding to either the expanding ($H > 0$) or contracting 
($H < 0$) phase.   As discussed in \cite{Pavlovic:2020sei}, it is natural that 
in cyclic models the value of the Ricci scalar would approach its maximum around 
the bounce and since $\dot{H} >0$, from \eqref{eq71} it follows that this 
maximum Ricci scalar value is positive, and moreover that  $\ddot{H}=0$  at the 
bounce.

In summary,   at the bounce we have $R= R_{\text{bounce}} > 0$, $H=0$, 
and $a=a_{\text{min}}$. The bounce is then followed by a phase in which $\dot{H} 
> 0$, $H>0$, $\dot{a}> 0$, and $\dot{R} < 0$.  Then  the Universe  enters the 
phase characterized by $\dot{H}< 0$ and approaches the turnaround point which 
is determined by $H=0$, $a=a_{\text{max}}$, and $R= R_{\text{turnaround}} < 0$, 
where the last condition follows from Eq. \eqref{eq71}. 

To obtain $g(a, H, R)$ in Eq. \eqref{eq_g}, we use Eqs. \eqref{gfe1}, 
\eqref{matterconsv}, \eqref{H1},
\eqref{H2}, and  \eqref{PDE}, where for simplicity we focus on the dust case. 
 Therefore, we obtain
\begin{align}
& R= 6 \dot{H}+12 H^2= -24 \pi G p_{DE}\nonumber\\
& = 3 \Bigg\{-8 \pi  G  \rho_{m} \left[\text{sech}\left(\frac{\beta  
H_0^2}{H^2}\right)-1\right] \nonumber \\
& +3 \beta 
   H_0^2 \text{shi}\left(\frac{H_0^2 \beta }{H^2}\right)-3 H^2 \left[\cosh 
\left(\frac{\beta 
   H_0^2}{H^2}\right)-1\right]+\Lambda \Bigg\}.
\end{align}  
Introducing the dimensionless quantities 
$
    E= \frac{H}{H_0}$ and $\mathcal{R}= \frac{R}{12 H_0^2}$,
 and using  $z$ as the independent variable, we extract the general 
system for $(a, E, \mathcal{R})$: 
\begin{align}
   &\frac{da}{dz}= -a^2, \label{eqAa}\\
   &\frac{dE}{dz}= -2\left(\mathcal{R} -  E^2\right) \frac{a}{E},\label{eqAb} \\
   & \frac{d\mathcal{R}}{dz}= -\frac{9 \beta {\Omega_m^{(0)}}^2 \tanh 
\left(\frac{\beta}{E^2}\right)
   \text{sech}^2\left(\frac{\beta}{E^2}\right)}{4 E^4 a^5}. \label{eqAc} 
\end{align}

Finally, in order to examine whether the above requirements are 
fulfilled in  the present scenario, we use  the best fit values $\beta=-0.011$ 
and $\Omega_m^{(0)}=0.283$ in \eqref{eqAa}, \eqref{eqAb} and \eqref{eqAc}, and 
we find 
    \begin{align}
   & \frac{da}{dz}= -a^2, \label{systAa}\\
   & \frac{dE}{dz}=- 2\left(\mathcal{R} -  E^2\right)\frac{a}{E},  
\label{systAb}
   \\
   & \frac{d\mathcal{R}}{dz}= -\frac{0.0019822 \tanh 
\left(\frac{0.011}{E^2}\right)
   \text{sech}^2\left(\frac{0.011}{E^2}\right)}{E^4 a^5}  \label{systAc}.
\end{align}
In the case of dust matter and $\Lambda\neq 0$. This system cannot satisfy the 
above requirements, and hence the present scenario cannot exhibit bounce and 
turnaround solutions.

 \subsection{Case II: $\Lambda = 0$}
  
 In the  case $\Lambda = 0$,  eq. \eqref{gfe2} becomes 
\begin{equation}
\frac{8 \pi  G  \rho_{m}}{3 H^2}  =- \frac{3 \pi  K \text{shi}\left(\frac{K \pi 
}{G
  H^2}\right)}{3 G H^2 } + \cosh \left(\frac{\pi  K}{G
  H^2}\right).
\end{equation}
\begin{figure} 
    \centering
   \includegraphics[scale=0.6]{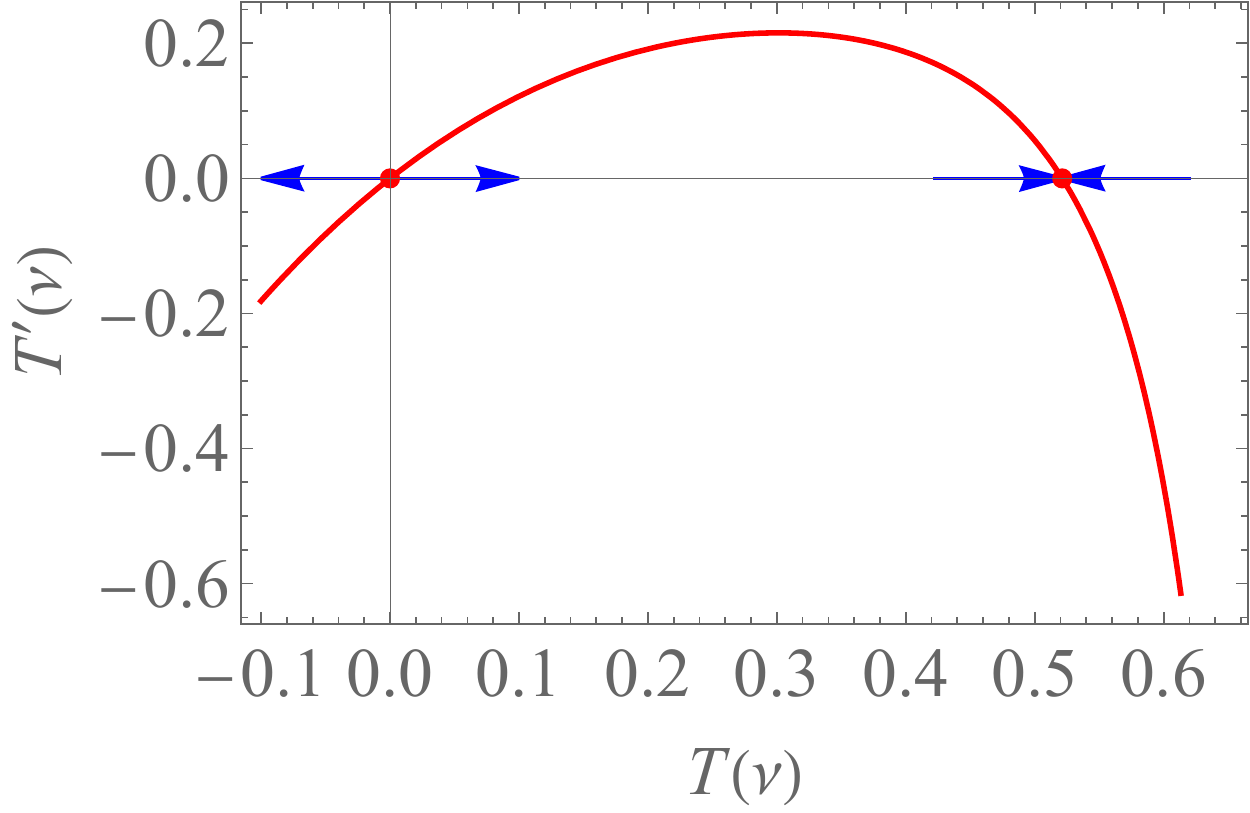}
    \caption{Phase-space diagram of the one-dimensional dynamical system 
\eqref{one-D} of 
Kaniadakis horizon entropy cosmology with $\Lambda=0$,  for dust matter and  for 
the best-fit value of Kaniadakis parameter obtained by the observational 
analysis, namely for $\beta=1.161$. The physical region is $0\leq T< 1$. The 
equilibrium point $T=0$ is unstable,
dominated by dark-energy, and de Sitter equilibrium point 
$T=T_c\approx 0.521$ is stable.}
    \label{F1D}
\end{figure}
This expression is used as a definition of $\rho_m$. 
If $\beta\neq0$, and re-scaling the time derivative $d/d\nu = (1-T)d/d\tau$, we 
obtain 
\begin{small}
\begin{align}
 & T^{\prime}(\nu)=\frac{3}{2} (1-T)^2 T \cosh \left(\frac{\beta  
T^2}{(1-T)^2}\right) -  \frac{3}{2}\beta   T^3 \text{shi}\left(\frac{T^2 \beta 
}{(1-T)^2}\right).\label{one-D}
\end{align}
\end{small}
The equilibrium points of \eqref{one-D} are $T=0$, which is unstable, and the  
equilibrium point 
$T=T_c$, where $T_c$ is a solution of the transcendental equation 
\begin{small}
$(1-T_c)^2 \cosh \left( {\beta  T_c^2}/{(1-T_c)^2}\right) - \beta   T_c^2 
\text{shi}\left( {T_c^2 \beta }/{(1-T_c)^2}\right)=0$
\end{small}, $0<T_c<1$, corresponding to de Sitter solution $a(t)\propto e^{H_0 
t \left(\frac{1}{T_c}-1\right)}$, is stable. In Fig. \ref{F1D} we depict a 
phase-space plot of the one-dimensional dynamical system \eqref{one-D} of 
Kaniadakis horizon entropy cosmology with $\Lambda=0$,  for dust matter, and the 
join value $\beta=1.161$.  Note that all orbits originate from the invariant 
subset $T=0$, classically related to the initial 
singularity with $H \rightarrow \infty$. The late-time attractor is 
$T=T_c\approx 0.521$, and it corresponds to de Sitter solution.

Finally, 
in order to examine whether the present scenario exhibits 
a bounce, we use   the best fit values $\beta=1.161$ and $\Omega_m^{(0)}=0.326$ 
in \eqref{eqAa}, \eqref{eqAb} and \eqref{eqAc} and we find 
    \begin{align}
   & \frac{da}{dz}= -a^2, \label{systBa}\\
   & \frac{dE}{dz}= -2\left(\mathcal{R} -  E^2\right) \frac{a}{E}, 
\label{systBb}\\
   & \frac{d\mathcal{R}}{dz}= \frac{0.277619 \tanh 
\left(\frac{1.161}{E^2}\right)
   \text{sech}^2\left(\frac{1.161}{E^2}\right)}{E^4 a^5}, \label{systBc}
\end{align}
in the case  of dust matter 
and $\Lambda=0$,
we deduce that the system cannot fulfill  the 
bounce requirements, and therefore it cannot exhibit bounce and 
turnaround solutions.

\section{Summary and discussion} \label{sec:Con}

The present work was devoted  to explore Kanadiakis horizon entropy cosmology 
which arises  from the application of the gravity-thermodynamics conjecture 
using the Kaniadakis modified entropy. The resulting  modified Friedmann 
equations   contain extra terms that constitute an effective dark energy 
sector. 
Moreover, we used data from     
Observational Hubble Data, Supernova Type Ia,  HII galaxies, Strong Lensing 
Systems, and  Baryon 
Acoustic Oscillations observations, and we applied  a Bayesian Markov 
Chain Monte Carlo analysis in order to construct the likelihood contours 
for the model parameters.

Regarding the Kaniadakis parameter $\beta$, we found that it is constrained  
around 0, namely  around the value in which standard Bekenstein-Hawking is 
recovered. Furthermore, the present matter density parameter $\Omega_m^{(0)}$ 
is 
consistent with the expected value from $\Lambda$CDM scenario, having a lower 
value for the $\Lambda\neq0$ case and a slightly higher value for the 
$\Lambda=0$ case.  

However, the interesting result comes from the   constraint 
on the normalized Hubble parameter $h$. In particular,  for $\Lambda \neq 0$ 
we extracted  $h=0.708^{+0.012}_{-0.011}$ while for $\Lambda=0$ we found
$h=0.715^{+0.012}_{-0.012}$. Thus, the obtained value of $H_0$ for $\Lambda 
\neq 
0$   deviates $2.67\sigma$ from the 
  Planck value  and $1.74\sigma$  from the SH0ES one, while in the 
 $\Lambda=0$ case the deviation is 
 $3\sigma$  from the Planck value and $1.36\sigma$ from the SH0ES one.
Additionally, in order to verify this result in an
independent way, we  performed the  $\mathbf{\mathbb{H}}0(z)$ 
diagnostic. Hence, our analysis reveals
 that   Kaniadakis horizon entropy cosmology is an interesting 
candidate to alleviate the  $H_0$ tension problem. This is one of the main 
results of the present work.

We proceeded by investigating the cosmographic parameters, namely the 
deceleration and jerk  ones, by using the data in order to  reconstruct them in 
the redshift region $0<z<2.5$. As we showed, the transition from 
deceleration to acceleration happens at $z_T=0.715^{+0.042}_{-0.041}$ for the 
$\Lambda \neq 0$ case and at   $z_T=0.652^{+0.032}_{-0.031}$ for the 
$\Lambda =0$ case, in agreement  within $1\sigma$ with that found in 
\cite{HerreraZamorano:2020rdh} for $\Lambda$CDM cosmology. 
Furthermore, 
we applied the AICc and  BIC  information criteria and we found that 
although AICc suggests that 
$\Lambda \neq 0$ model and $\Lambda$CDM are statistically equivalent in the 
joint analysis, BIC indicates that there is a strong evidence against   
the candidate model. Lastly, applying the  DIC criterion we found that 
the $\Lambda \neq 0$ case and 
$\Lambda$CDM are statistical equivalent for BAO, they have a moderate tension 
for OHD and SLS, and a strong tension for HIIG and SNIa datasets, while  the 
$\Lambda \neq 0$ case and $\Lambda$CDM are statistical equivalent for 
all datasets.

Finally, we performed a detailed dynamical-system analysis, providing a 
general description of the phase-space of all possible solutions of the 
system, their equilibrium points and stability, as well as the late-time 
asymptotic behavior. As we showed, the Universe past attractor  is the 
matter-dominated epoch, while at  late times the Universe results in the 
dark-energy-dominated solution, for both $\Lambda=0$, and $\Lambda \neq0$ cases.
Moreover,  we showed that the scenario accepts  
heteroclinic sequences, but it cannot lead to bounce and turnaround 
solutions.

In summary, the scenario of Kaniadakis horizon entropy cosmology exhibits very 
interesting phenomenology and is in agreement with observational behavior. 
Hence, it can be an interesting candidate for the description of Nature.

\section*{Acknowledgments}  
We thank the anonymous referee for thoughtful remarks and suggestions. A.H.A. thanks to the PRODEP project, Mexico for resources 
and financial support and thanks also to the support from Luis Aguilar, 
Alejandro de Le\'on, Carlos Flores, and Jair Garc\'ia of the Laboratorio 
Nacional de Visualizaci\'on Cient\'ifica Avanzada.
G.L. was funded by  Agencia Nacional de Investigaci\'on y Desarrollo - ANID for 
financial support through the program FONDECYT Iniciaci\'on grant no. 11180126 
and by Vicerrectoría de Investigación y Desarrollo Tecnológico at UCN. J.M. acknowledges the support from ANID project Basal AFB-170002 and ANID REDES 
190147. M.A.G.-A. acknowledges support from Universidad Iberoamericana that support with the SNI grant, ANID REDES (190147), C\'atedra Marcos Moshinsky and Instituto Avanzado 
de Cosmolog\'ia (IAC).  V.M. acknowledges support 
from Centro de Astrof\'{\i}sica de Valpara\'{i}so and ANID REDES 190147. This 
work is partially supported by the Ministry of Education and Science of the Republic of Kazakhstan, Grant AP08856912. A.D. Millano was supported by Agencia Nacional de Investigación y Desarrollo - ANID-Subdirección de Capital Humano/Doctorado
Nacional/año 2020- folio 21200837 and by Vicerrectoría de Investigación y Desarrollo Tecnológico at UCN.

\section*{Data Availability}
The data underlying this article were cited in Section \ref{subsec:data}.
 


\bibliographystyle{mnras}
\bibliography{main}

\bsp	
\label{lastpage}
\end{document}